\documentclass[letterpaper]{article} %
\usepackage{aaai25}  %
\usepackage{times}  %
\usepackage{helvet}  %
\usepackage{courier}  %
\usepackage[hyphens]{url}  %
\usepackage{graphicx} %
\usepackage{natbib}  %
\usepackage{caption} %
\usepackage{algorithm}
\usepackage{tikz,graphicx,xcolor}
\usepackage{gensymb}
\usepackage{textcomp}
\usepackage{float,dblfloatfix}
\usepackage{verbatim,array,enumitem}
\usepackage{algorithm,algpseudocode,breqn,amsmath}
\usepackage{adjustbox,diagbox,threeparttable,tablefootnote,colortbl}
\usepackage{booktabs,tabularx,multirow,multicol,hhline}
\usepackage{caption,subcaption}
\PassOptionsToPackage{hyphens}{url} 
\usepackage[capitalise,noabbrev]{cleveref} 
\crefname{appsec}{Appendix}{Appendices} 
\usepackage[misc]{ifsym}
\usepackage[utf8]{inputenc}
\usepackage{booktabs}
\usepackage{lipsum}
\usepackage{tabularray}
\usepackage{longtable}
\usepackage{mdframed}
\usepackage{tcolorbox}
\usepackage{multicol}

\usepackage{amssymb}

\usepackage{newfloat}
\usepackage{listings}
\DeclareCaptionStyle{ruled}{labelfont=normalfont,labelsep=colon,strut=off} %
\floatstyle{ruled}
\newfloat{listing}{tb}{lst}{}
\floatname{listing}{Listing}
\usepackage{xcolor}
\newcommand{\answerYes}[1]{\textcolor{blue}{#1}} 
\newcommand{\answerNo}[1]{\textcolor{teal}{#1}} 
\newcommand{\answerNA}[1]{\textcolor{gray}{#1}}

\title{\textit{Promptly Yours?} A Human Subject Study on \\Prompt Inference in AI-Generated Art}

\author{
\textnormal{Khoi Trinh\textsuperscript{\rm 1}}, 
\textnormal{Joseph Spracklen\textsuperscript{\rm 2}}, 
\textnormal{Raveen Wijewickrama\textsuperscript{\rm 2}}, \\
\textnormal{Bimal Viswanath\textsuperscript{\rm 3}}, 
\textnormal{Murtuza Jadliwala\textsuperscript{\rm 2}}, 
\textnormal{Anindya Maiti\textsuperscript{\rm 1}} \\
\textnormal{\textit{khoitrinh@ou.edu, joseph.spracklen@my.utsa.edu, raveen.wijewickrama@utsa.edu, \\vbimal@vt.edu, murtuza.jadliwala@utsa.edu, am@ou.edu}}
}

\affiliations {
    \textsuperscript{\rm 1} University of Oklahoma\\
    \textsuperscript{\rm 2} University of Texas at San Antonio\\
    \textsuperscript{\rm 3} Virginia Tech
}

\usepackage{bibentry}

\begin{document}

\maketitle

\begin{abstract}

The emerging field of AI-generated art has witnessed the rise of prompt marketplaces, where creators can purchase, sell, or share prompts to generate unique artworks. These marketplaces often assert ownership over prompts, claiming them as intellectual property. This paper investigates whether concealed prompts sold on prompt marketplaces can be considered as secure intellectual property, given that humans and AI tools may be able to approximately infer the prompts based on publicly advertised sample images accompanying each prompt on sale. 
Specifically, our survey aims to assess (i) how accurately can humans infer the original prompt solely by examining an AI-generated image, with the goal of generating images similar to the original image, and (ii) the possibility of improving upon individual human and AI prompt inferences by crafting human-AI combined prompts with the help of a large language model. 
Although previous research has explored the use of AI and machine learning for prompt inference (and also to protect against it), we are the first to include humans in the loop. 
Our findings indicate that while humans and human-AI collaborations can infer prompts and generate similar images with high accuracy, they are not as successful as using the original prompt.

\end{abstract}

\section{Introduction}
\label{sec:introduction}

Artificial Intelligence (AI) has made remarkable strides in the domain of creative and artistic expression, enabling an easy and automated process for everyone to generate visually captivating and conceptually intriguing art work. Central to this creation process of AI-generated art and images are deep learning based text-to-image (\texttt{txt2img}) models for image generation that utilize text prompts as input (instructions) from users to generate unique and diverse image/art outputs. 
Some of the most popular open-source or commercially-available examples of such \texttt{txt2img} models include Midjourney~\cite{midjourney}, DALL-E 2~\cite{ramesh2021zero}, Stable Diffusion~\cite{rombach2022high} and GLIDE~\cite{nichol2022glide}. 

At a high level, these models have two main components -- a Language Model (e.g., CLIP\footnote{\url{https://openai.com/research/clip}}) and a Generative image model (e.g., Stable Diffusion~\cite{rombach2022high}). The language model converts a given text prompt to a latent representation, which is then used to condition the generative image model to produce an image that captures the prompt description.
Furthermore, these models are trained on vast datasets of text-image pairs, allowing them to understand and render complex visual concepts from textual descriptions with remarkable accuracy and creativity. For example, Stable Diffusion was trained on the publicly available LAION-5B dataset, containing 5 billion image-caption pairs, derived from data scraped from the Internet.

Text prompts serve as critical input instructions to \texttt{txt2img} models for generating high-quality text-conditioned images and it is non-trivial to deduce an appropriate prompt for the desired image, often requiring creativity and trial-and-error~\cite{wang2023review}. This has resulted in the emergence of new prompt engineering jobs and prompt marketplaces for AI-generated art, where prompt engineers, artists, and enthusiasts can exchange and sell prompts that can generate custom high-quality art.
Given the importance of selecting the right prompt for generating a desired image and the non-triviality in determining one, these text prompts are often treated as protected information by their creators. Prompt marketplaces often claim intellectual property rights over the prompts, asserting that they are valuable and original creations worthy of legal protection~\cite{promptbasetos,promptrrtos}. Given the protected status of (input) text prompts, two research questions arise that are largely unexplored thus far: (i) how accurately can humans infer the input text prompt by just viewing the (AI-generated) image generated from that prompt? and (ii) can AI tools assist humans in more accurately inferring text prompts of a target (AI-generated) image?

To address these research questions, our study employs a human subject survey to assess how accurately users (participants) can infer prompts by just visually examining the AI-generated images. 
Our survey results provides valuable insights on how users' prediction (of prompts) performance, measured using well-defined metrics, varies with different prompt-related attributes and varying user demography and backgrounds. 
After combining responses from the survey (taken by human subject participants) with responses from an AI-based prompt inference model, we further re-evaluate and compare the overall inference accuracy in this human-AI collaborative setting. 
Our results show that although both human and combined human-AI efforts can accurately infer prompts and recreate images to a great extent, they fall short of the effectiveness achieved with the original prompts. Consequently, marketplaces for selling prompts and creators who offer prompts for AI-generated art can continue to maintain a viable business model.

\section{Related Work and Research Goals}
\label{sec:related}

\subsection{Prompt Inference in AI Art}
Previous research in the literature has primarily explored AI/ML-based inference techniques to deduce (infer) prompts and has also proposed AI/ML-based methods to safeguard against such kind of prompt inference threats. Several prior works have employed machine learning algorithms to reverse-engineer the prompts used in the creation of artworks~\cite{clipinterrogator,shen2023prompt,wu2022membership,li2022blip}, highlighting the potential vulnerability of prompt concealment. On the other hand, efforts have also been made to develop protective measures to secure the prompts from unauthorized access or replication~\cite{shen2023prompt,struppek2022rickrolling,zhai2023text}. These efforts involve strategies to thwart prompt inference, backdoor injections, and data poisoning, implementing rigorous dataset inspections, and employing anomaly detection. 
However, if humans are able to infer prompts with a high degree of accuracy, the effectiveness of these protective measures against prompt inference may be called into question.
Furthermore, it is possible that AI-assisted prompt inference and prompt inference by humans can be effectively combined to significantly improve the overall inference accuracy, and needs to be further studied. While there are recent studies investigating the effectiveness of human-AI collaboration leading to more creative artworks~\cite{lyu2022communication,brade2023promptify}, currently we have little understanding of how a similar human-AI collaboration may work towards prompt inference of AI-generated art.

\subsection{Challenges in Prompt Inference}

\begin{figure}[t]
\centering
\includegraphics[width=0.99\linewidth]{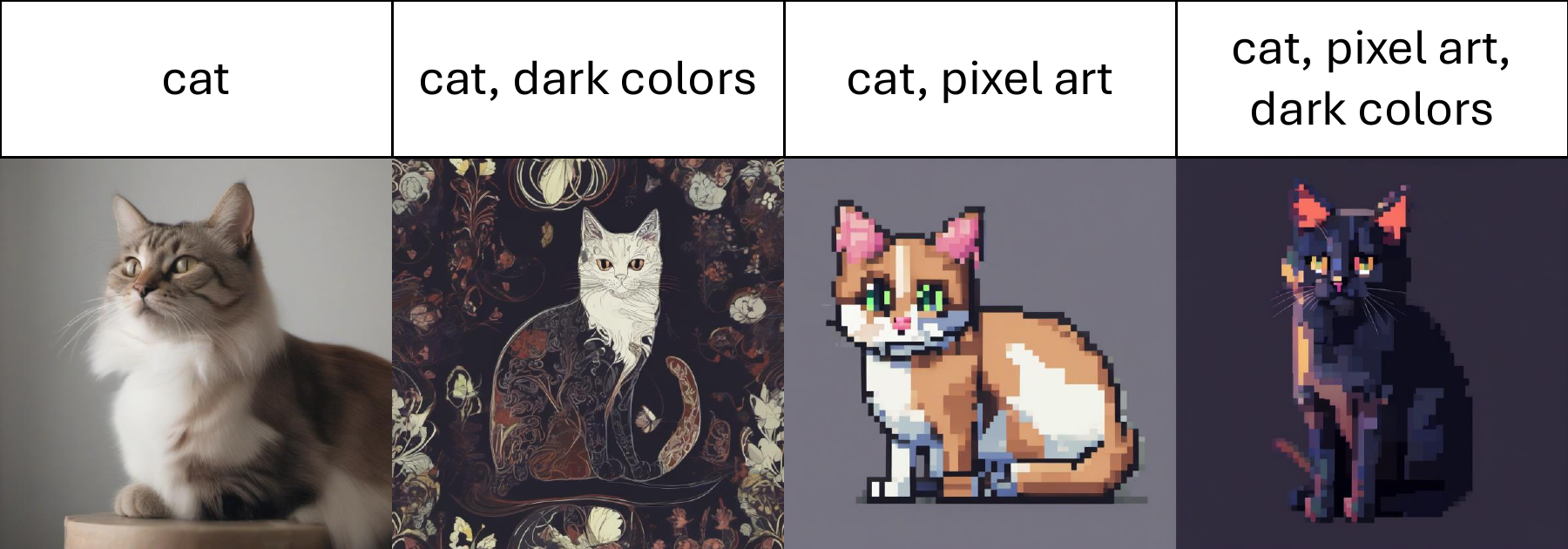}
\caption{Image generations using SDXL with prompts containing the same subject (cat) and different combinations of two modifiers (pixel art and dark colors).}
\label{fig:sub-mod-example}
\end{figure}

Inferring prompts of AI-generated images is a complex task that both humans and AI models~\cite{clipinterrogator,img2prompt,shen2023prompt} can find challenging due to the complexities in visual content and the subtleties of language. \cref{fig:sub-mod-example} exemplifies this by illustrating how varying the modifiers (in this case, ``pixel art" and ``dark colors") for a consistent subject (a cat) can lead to vastly different visual outcomes when combined with different modifiers. 
Even the addition of a single modifier can dramatically alter the resulting image, demonstrating how each element in a prompt contributes to the generated image. 
Conversely, omitting a crucial modifier could lead to a visual representation that misses the depth or context intended.
When humans try to deduce the prompts for AI-generated images, they often rely on their subjective interpretation and understanding of the visual content, which can lead to varied conclusions on the subjects and modifiers used in the image generation. 
The understanding of how humans may infer prompts, specifically the subjects and modifiers used in AI-generated art, with or without the aid of AI-based prompt inference tools, remains an unexplored research area and is the focus of this work.

\subsection{Research Goals}

Given the above intricacies in human and AI prompt inferences, it is not trivial to assess whether concealed prompts sold on prompt marketplaces can be considered as secure intellectual property, or are they vulnerable to prompt inference and replication. Moreover, there is a lack of a clearly defined measurement based on which a prompt inference can be deemed successful, especially for human prompt inference. To address this gap, our research goals are as follows:

\begin{itemize}[leftmargin=*]
    \item Determining the accuracy with which individuals can infer the original prompts from images created by a \texttt{txt2img} model, aiming to reproduce similar images. 
    This is accomplished by designing and conducting a comprehensive human participant survey as outlined in  \cref{sec:survey} and \cref{sec:exp-setup}. Additionally, exploring whether a participant with an art background (e.g. an art major in college, a graphic designer) would have an advantage over one without such a background.
    Evaluation is to be based on similarity between the original image and the images reproduced using the inferred prompt.
    \item Determining any improvement in prompt inference accuracy when human inferred prompts are combined with AI inferred ones, aiming to further improve the similarity of the reproduced images to the original images. This is achieved by integrating the survey responses of the human participants and AI inferred prompts as discussed in \cref{sec:human-ai-expsetup}. Evaluation is to be based on similarity between the original image and the images reproduced using the combined prompt. 
    \item Establishing robust thresholds of metrics for measuring the success of both standalone and combined human-AI inferences, as detailed in \cref{sec:success-scores-expsetup}, ensuring a clear framework for assessment.
\end{itemize}

\section{Survey Design and Participants}
\label{sec:survey}

\subsection{Prompt and Image Datasets}
\label{sec:dataset}

The images presented to the participants were generated from two distinct types of prompts: \textit{Controlled} and \textit{Uncontrolled}. Controlled prompts serve as a baseline to gauge participants' ability 
to recognize common subjects and modifiers, %
allowing us to assess other variables associated with prompt inference, such as any differences between \texttt{txt2img} models and demographic-based differences. %
Conversely, uncontrolled prompts feature a much more diverse mix of subjects and modifiers and is used to construct more comprehensive and challenging inference tasks for participants.

\subsubsection{Controlled Dataset.}
\label{sec:controlled-prompts}

The controlled prompts set was constructed through an analysis of common subjects discovered on the dynamically changing homepage of Lexica\footnote{\url{https://lexica.art}}, a popular prompt and AI art sharing platform. %
We selected Lexica for this data gathering task as they have web crawling-friendly terms of service. 
Our \textit{Selenium} script identified 100579 different images and accompanying prompts by their specific HTML elements. Subsequent processing with \textit{spaCy}, a Natural Language Processing (NLP) framework, enabled us to distill the subjects from these prompts. The five most frequently occurring subjects identified were \textit{man}, \textit{woman}, \textit{astronaut}, \textit{cat}, and \textit{robot}. 
In parallel with subject selection, we curated a set of modifiers by referencing a Midjourney styles and keywords repository~\cite{midjourneymodifiers}. The modifiers spanned various categories such as themes/genres, lighting, drawing and art medium, perspective, emotion/mood, colors and palettes, geography and culture, rendering/shading style, culminating in a list of 121 modifiers (details in \cref{appendix-modifiers}). 
These subjects and modifiers were then randomly sampled and combined, ranging from one to five modifiers per subject, to formulate 100 controlled prompts, 20 per subject (\cref{appendix-prompts}). These 100 prompts were then used to generated images for Parts~I and III of our survey (\ref{sec:survey}), 25 for each of the four selected \texttt{txt2img} models.

\subsubsection{Uncontrolled Dataset.}
\label{sec:uncontrolled-prompts}

The uncontrolled prompts set was constructed through a random sampling of 100 whole/complete prompts discovered on PromptHero (a platform for discovering and sharing AI-generated art and text prompts).
There were no restrictions on what subject and modifiers appeared in these prompts, as long as they were non-explicit in nature. 
These 100 prompts were then used to generate images for Part~II of our survey, 25 for each of the four selected \texttt{txt2img} models.

\subsection{Analyzed \texttt{txt2img} Models}
\label{sec:txt2img}

Our study focuses on four popular \texttt{txt2img} models, including two base models and two fine-tuned models:

\begin{itemize}[leftmargin=*]
\item \textbf{MidJourney v5.0}~\cite{midjourney}.
\item \textbf{Stable Diffusion XL} (SDXL)~\cite{podell2023sdxl,rombach2022high}.
\item \textbf{DreamShaper XL}~\cite{podell2023sdxl}.
\item \textbf{Realistic Vision v5}~\cite{rv5}.
\end{itemize}

At the time we began this study in late 2023, these models had the highest number of prompts being shared or sold on platforms such as PromptHero, Promptrr.io, Prompti AI, PromptBase, and CivitAI. More details on these four models and how they were employed in our study can be found in~\cref{appendix:txt2img-models-details}.

Besides the above, there are other popular \texttt{txt2img} models which we also attempted to include in our study, but could not for various reasons. For instance, OpenAI's popular DALL-E model utilizes a modified version of the GPT model by adapting its transformer-based architecture for image generation~\cite{ramesh2021zero}. However, DALL-E was omitted from our study because we were unable to lock in a specific version of the model for image creation. We were concerned that DALL-E's updates during the time-frame of our study could affect the consistency of the images produced initially and those generated later for comparison.

For our user study to assess the ability of humans to infer prompts from AI-generated images, we design and implement a custom survey tool. This tool dynamically loads and displays images (and its metadata) together with related question(s), and can capture participant responses to these questions (see \cref{fig:cat-grading,appendix-p1,appendix-p2,appendix-p3}). 
We collected responses from 230 participants using this custom tool, of which 59 were recruited from two public university campuses (in the US), while the remaining 171 participants were recruited via the Amazon Mechanical Turk (MTurk) platform. 
This study has been approved by the Institutional Review Boards (IRBs) of the institutions involved.
Before completing the main survey task, participants completed a preliminary questionnaire (\cref{fig:demo-surveyform} in \cref{appendix:survey-details}) that requested demographic information and self-reported familiarity with generative AI tools. Below, we first summarize findings from the preliminary survey, followed by a detailed description of the main survey tasks completed by the participants.

\subsection{Pre-Survey: Demographics}

30.9\% of the participants in our study are in the age group of 35-44 years, followed by 26.5\% in 25-34 years and 20.4\% in 18-24 years. A small number of remaining participants are 45 years or older.
\cref{fig:age-distribution} and \cref{fig:gender-distribution} (in \cref{appendix:presurvey-details}) illustrate the age and gender distribution of our participants, respectively. 
 Most of our participants are male (59.1\%), with the remaining females (38. 7\%) and those who identify as other gender categories (2.2\%). 
15.65\% participants reported having an arts background or education, indicating a certain level of expertise or familiarity with creative fields, while the remaining 84. 35\% did not have such a background, representing a broader range of occupations and experiences. %

\subsection{Pre-Survey: Familiarity with Generative AI}

\begin{figure}[t]
\centering
\includegraphics[width=0.99\linewidth]{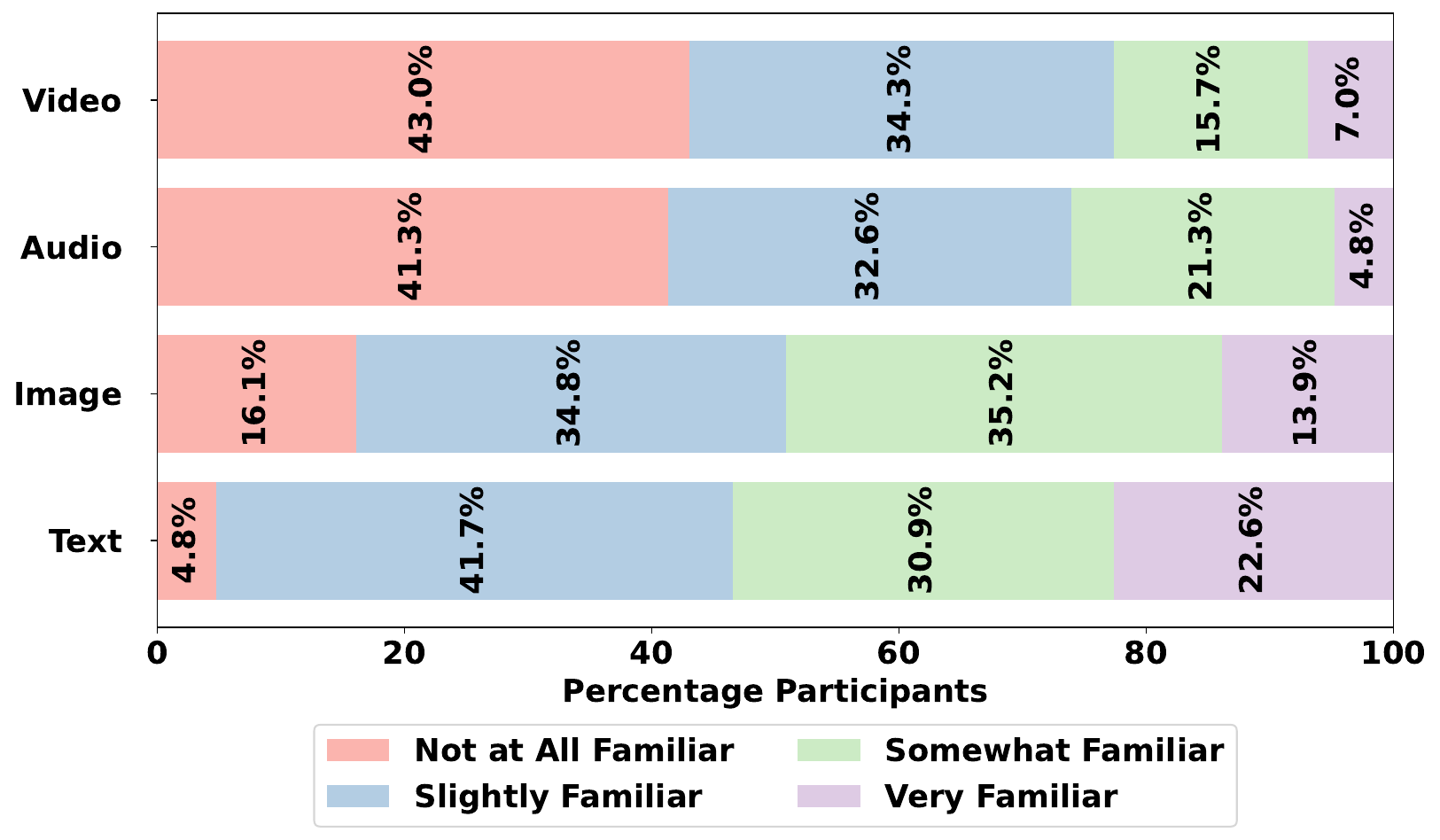}
\caption{Participants' familiarity levels with different generative AI tools.} %
\label{fig:ai-tool-familiarity}
\end{figure}

We separated the level of familiarity of our participants with various generative AI tools into four distinct categories, namely, familiarity with the text, image, audio, and video generative AI tools (\cref{fig:ai-tool-familiarity}). 
The predominant level of familiarity for text and audio tools is ``\emph{Slightly Familiar},'' (41.7\% for text and 32.6\% for audio), suggesting a moderate acquaintance with these technologies. 
For image generation tools, the results indicate an evenly spread level of familiarity, with both ``\emph{Slightly Familiar}'' and ``\emph{Somewhat Familiar}'' categories capturing the largest proportions at roughly 35.0\% each. For video generation tools, both ``\emph{Not At All Familiar}'' and ``\emph{Slightly Familiar}'' categories capture the largest proportions at 43.0\% and 34.3\%, respectively. %
``\emph{Very Familiar}'' holds smaller shares across video and audio tools (4.8\% for audio and 7.0\% for video), while for image and text generation tools ``\emph{Not at All Familiar}'' remains the least represented (4.8\% for text and 16.1\% for image). This distribution demonstrates that our participants are moderately familiar with a broad range of generative AI tools.

\subsection{Parts~I-III: Prompt Inference}
Our pool of participants, both recruited on campus and through MechTurk, participated in the main survey, comprised of three sequentially ordered parts (or phases), where 
in each part participants were tasked with either selecting or typing-in the most appropriate prompt (comprising of a subject and modifiers) for a series of AI-generated images.
In Parts~I and III, a selection of 100 images using a controlled set of subjects and modifiers, each methodically produced using the four image generation models outlined in \cref{sec:txt2img}, were displayed to the participants. Each of these models contributed an equal share of images to the controlled dataset. 
Consistent exposure to all subjects and modifiers in the controlled set was maintained by appropriately varying the subset of images each of the participants were exposed to.

In contrast, Part~II expanded the scope of inference by using 100 images from an uncontrolled dataset. These images were generated from popular prompts found on online prompt sharing websites, offering a wide array of subjects and modifiers. This uncontrolled set also used the same four image generation models. Detailed information on the subjects and modifiers selected for the controlled dataset and 
the composition of the uncontrolled dataset can be found in \cref{sec:dataset}. %

\subsubsection{Part~I: Subject and Modifier Selection in Controlled Dataset.}
\label{sec:survey-p1}
In this part, participants were presented with a sequence of five AI-generated images with controlled subjects and modifiers (i.e, from the controlled dataset). For each displayed image, they were required to first select the correct subject from the options provided. Next, they had to choose the correct modifier(s) from the given options (between 1 to 5 checkbox options), with the possibility of selecting multiple modifiers if preferred. For each question, there will be one correct subject, with the rest of the options filled in randomly; at least one and at most five correct modifiers. For questions where there are less than five correct modifiers, the remaining modifiers are filled in randomly. \cref{fig:p1} shows an example of a question for this part.
The goal of this task was to measure the participants' ability to discern and match the given images to their corresponding prompts accurately.

\subsubsection{Part~II: Prompt Creation in Uncontrolled Dataset.}
\label{sec:survey-p2}
This second part challenged participants with another five AI-generated images, but this time the prompts were uncontrolled, i.e., no pre-selected options were provided for both subjects and modifiers. In this task, for each image 
participants had to type what they believed to be the most fitting subject and modifier(s) as a complete sentence. \cref{fig:p2} (in (in \cref{appendix-p2}) shows an example of a question for this part. This open-ended task assessed their creative inference abilities and how they translated their interpretation of the images into descriptive prompts.

\subsubsection{Part~III: Prompt Creation in Controlled Dataset.}
\label{sec:survey-p3}
In this final task, participants were shown yet another set of five AI-generated images. Similar to Part~II, they were instructed to generate sentence-like prompts for these images. However, in Part~III the images were again based on controlled subjects and modifiers (i.e., from the controlled dataset). \cref{fig:p3} (in \cref{appendix-p3}) shows an example of a question for this part. The responses of the participants were then compared with the actual prompts and images displayed to them, providing a measure of how well the participants could infer and replicate the structured prompts that were initially used to generate the images displayed. %

\subsection{Part~IV: Rating Similarity}
\label{sec:survey-p4}
In addition to parts I through III, a subset (all of our 171 MTurk participants) were asked to take an additional part. In this last part (Part~IV) of the study, participants rated the similarity between pairs of images, \textit{one shown to prior participants and one generated using their (prior participants') prompts}. %
Specifically, participants were presented with pairs of images, similar to the ones depicted in \cref{fig:cat-grading} (in \cref{appendix-p4}). Each pair consisted of an original image from Part~III and a newly generated image based on a prompt submitted by one of the first 25 (on campus) participants. The participants' task for each image pair was to assess and rate the similarity between the two images. They were provided with a Likert scale ranging from ``Not at All Similar'' to ``Very Similar''. This allowed participants to express their perceived degree of similarity, taking into account factors such as color scheme, composition, and subject representation. After making their selection, they would proceed to the next comparison until they had evaluated all five pairs assigned to them.

\subsection{Survey Responses}

In total, 230 on campus and MTurk
participants took our survey between December 2023 and March 2024.
The responses collected consists of subject and modifier selections from a fixed set of options in Part~I, and whole prompts in Parts~II and III of the survey. 
After completion of the survey and elimination of invalid responses, we recorded 1078 responses for Part~I, 1145 responses for Part~II, and 1141 responses for Part~III.
In addition, the participants who were selected for Part~IV of our study gave us 855 responses. 
In Parts~II and III, responses that were incoherent or irrelevant, such as the failure to specify at least one subject or modifier, were also excluded.

\section{Experimental Setup}
\label{sec:exp-setup}

\subsection{Metrics}
\label{sec:metrics}

Next, we define the metrics that we employ to assess the prompt inference accuracy of the survey participants in our experiments. By means of these metrics, we aim to compare and compute the discrepancy between the original prompts and participants' responses (inferred prompts), and between the original image (displayed to the participants) and images generated from their (inferred) prompts. A detailed description of these metrics can be found in~\cref{appendix:metrics}. 
These metrics provide a comprehensive assessment of both objective (e.g., image hash, perceptual similarity) and subjective (e.g., surveyed similarity ratings) aspects, ensuring a well-rounded evaluation of the participants' performance in the study. 

\begin{itemize}[leftmargin=*]
\item \textbf{MSQ Scores (Survey Part~I).} 
Score between 0 (no correct selection) to 2 (all correct selections) for each question to evaluate multiple selection questions (MSQs). 

\item \textbf{Image Hash\footnote{\url{https://pypi.org/project/ImageHash/}}.}
Hamming distance between the hashes of two images. A lower image hash score (difference) implies that the two images are similar, and vice versa.

\item \textbf{Perceptual Similarity~\cite{zhang2018perceptual}.}
A lower perceptual similarity score implies that the two images are similar, and vice versa.

\item \textbf{Image Embedding Similarity (CLIP Score)~\cite{wang2023exploring}.} 
CLIP score between an image and a prompt (text) where a higher score indicates greater relevance or similarity.
We employ two state-of-the-art CLIP models in our evaluation, OpenAI's ViT-L/14 Transformer\footnote{\url{https://huggingface.co/openai/clip-vit-large-patch14}} (L14) and ViT-B/32 Transformer\footnote{\url{https://huggingface.co/openai/clip-vit-base-patch32}} (B32).

\item \textbf{Text Embedding Similarity (Semantic Similarity)~\cite{reimers2019sentence}.} 
A higher semantic similarity implies the two prompts are similar, and vice versa. 

\item \textbf{Surveyed Similarity Rating (from Survey Part~IV).} 
Participants in Part~IV of the study rated pairs of images on a Likert scale consisting of ``Not at All Similar'', ``Slightly Similar'', ``Somewhat Similar'', and ``Very Similar.''

\end{itemize}

\subsection{Quantifying Successful Inferences}
\label{sec:success-scores-expsetup}

Given our study's focus on assessing the accuracy of human prompt inference is ultimately in understanding their ability to generate images resembling those presented to them, we now establish a quantifiable measure of \textit{success threshold} based on the above metrics, specifically image hash, perceptual similarity, and CLIP scores. 
Considering the inherent variability in \texttt{txt2img} generations for identical prompts, achieving perfect scores (1 for CLIP score, 0 for image hash, and perceptual similarity) is implausible even with completely accurate prompt inference, as illustrated by \cref{fig:cat-a,fig:cat-b} scores. 

Therefore, a more appropriate threshold for gauging successful inference involves analyzing the range of scores for images generated from the same prompt across different instances. 
Accordingly, to determine an effective \textit{success threshold}, we assessed the scores generated from identical prompts across multiple image creations, using our 100 controlled dataset prompts (\cref{appendix-prompts}). For each of these prompts, we generated two different images using SDXL, and calculated the image hash, perceptual similarity, and B32 and L14 CLIP scores.
After calculating the averages of these scores, we determined the success thresholds, denoted by $\theta$, as follows: $\theta_{\text{hash}} = 26.83$ for image hash, $\theta_{\text{ps}} = 0.575$ for perceptual similarity, and $\theta_{\text{B32}} = 0.875$ and $\theta_{\text{L14}} = 0.851$ for B32 and L14 CLIP scores, respectively.
Participants whose inferred prompts are able to generate images close to or better than these thresholds can be deemed successful in their inference.

\subsection{Human-AI Combined Inference}
\label{sec:human-ai-expsetup}

We now detail our experimentation on the effectiveness of combining human inferred prompts with those produced by AI models towards accurately recreating AI-generated art. Specifically, we utilize prompt responses collected in Part~III of our survey, which involved controlled dataset prompts, and AI-generated prompts obtained through the CLIP Interrogator, a model that analyzes an image and generates descriptive text prompts by leveraging OpenAI’s CLIP model to match images and text representations \cite{clipinterrogator}.
These human and AI prompt pairs are then consolidated into one combined prompt for comparative evaluation against only human inferred prompt, CLIP Interrogator prompt, and the success threshold, in \cref{sec:colab-eval}.

To construct a combined prompt from each pair of human prompt and corresponding CLIP Interrogator prompt, we employ a large language model such as GPT-4 \cite{openai2023gpt}, and instruct it to merge the two prompts into a succinct new prompt of up to 25 words without adding any extraneous information. An example of this process is illustrated below, where the first prompt was from a participant and the second prompt was generated by CLIP Interrogator:

\begin{tcolorbox}%
\small
\textbf{Instruction}: Combine these two prompts into a new prompt of 25 words, without any extra information added:\\
1. man dressed in steampunk in a steampunk factory\\
2. a man in a steampunk suit and top hat standing in front of a giant clock with gears, Bastien L. Deharme, steampunk, a character portrait, fantasy art

\textbf{GPT-4 Response}: a man in a steampunk suit and top hat stands in a factory, surrounded by giant clocks and gears, embodying a fantasy art portrait.
\end{tcolorbox}

The combined prompts were capped at 25 words to prevent GPT-4 from inadvertently introducing additional keywords when given unrestricted length. Employing a large language model was considered more suitable than merely concatenating the two prompts to avoid repeating subjects and modifiers. Such repetitions could skew the \texttt{txt2img} generations by inadvertently emphasizing repeated subjects and modifiers over others.

\section{Results and Analysis}
\label{sec:results}

\subsection{Subject and Modifier Selection Evaluation}
\label{sec:mcq-eval}

\begin{figure}[t]
\centering
\begin{subfigure}[b]{0.48\linewidth}
    \centering
    \includegraphics[width=\textwidth]{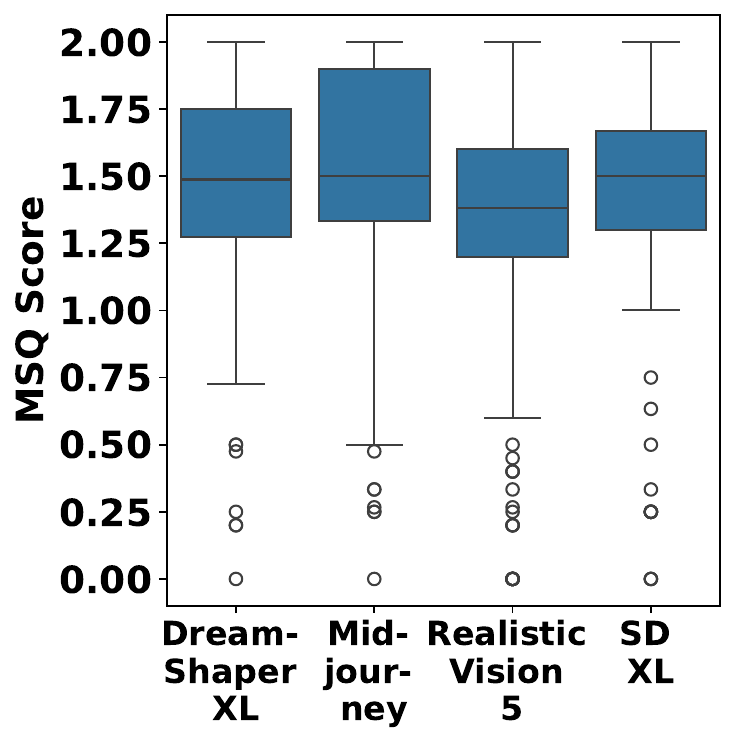}
    \caption{}
    \label{fig:mcq-model}
\end{subfigure}
\hfill
\begin{subfigure}[b]{0.48\linewidth}
    \centering
    \includegraphics[width=\textwidth]{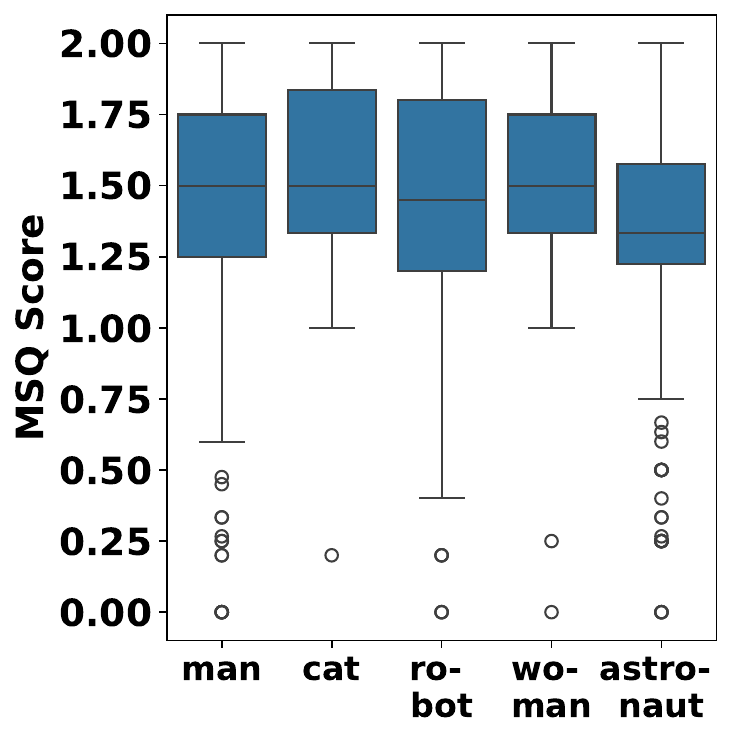}
    \caption{}
    \label{fig:mcq-subject}
\end{subfigure}
\hfill
\begin{subfigure}[b]{0.48\linewidth}
    \centering
    \includegraphics[width=\textwidth]{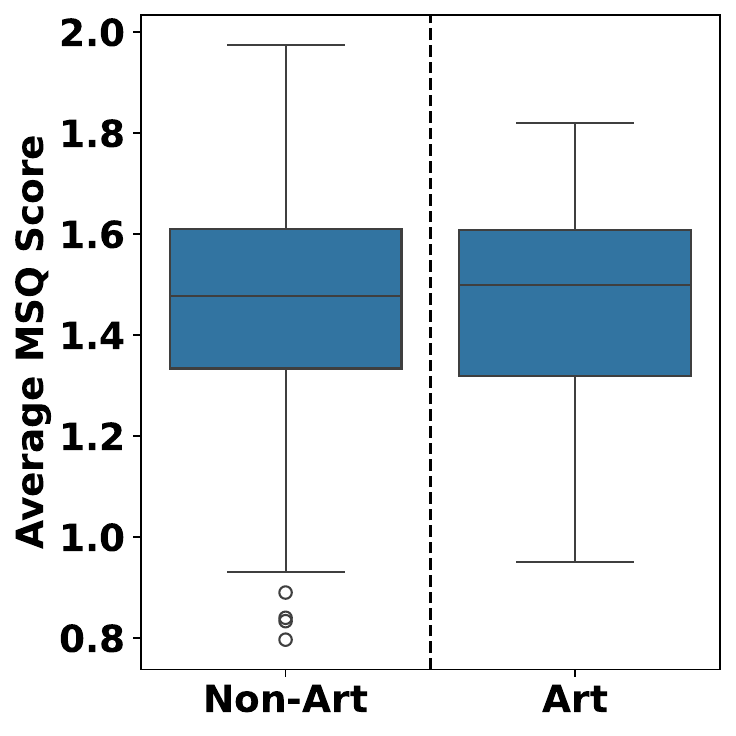}
    \caption{}
    \label{fig:mcq-art-bg}
\end{subfigure}
\hfill
\begin{subfigure}[b]{0.48\linewidth}
    \centering
    \includegraphics[width=\textwidth]{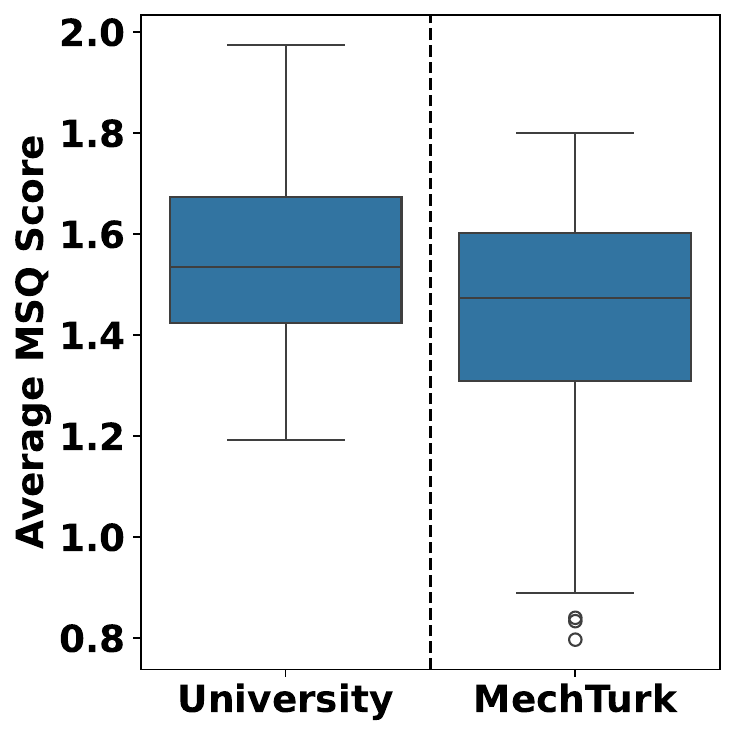}
    \caption{}
    \label{fig:mcq-mturk}
\end{subfigure}
\caption{Analysis of multiple answer question (MSQ) responses from Part~I, aimed at identifying disparities in (a) four distinct \texttt{txt2img} models, (b) various subjects depicted in the images, (c) the impact of participants' arts background, and (d) variations attributable to recruitment sources.}
\label{fig:mcq-eval}
\end{figure}

In \Cref{fig:mcq-eval}, the $y$-axis illustrates the Part I MSQ score distributions over the specified categories (models, subjects, and demographic), reflecting participants' ability to match AI-generated images with their corresponding subjects and modifiers where a higher score indicate a more accurate prompt matching. %
As seen in \cref{fig:mcq-model}, Midjourney-generated images showed a relatively high score concentration towards the upper end, with a median score of 1.5 and an interquartile range (IQR) ranging from 1.33 to 1.9. 
Realistic Vision 5's generations demonstrated a median of 1.38 and an IQR from 1.2 to 1.6. DreamShaper XL's images displayed a median of 1.49 and an IQR from 1.28 to 1.75, suggesting slightly more consistency than Realistic Vision 5. 
Stable Diffusion XL matched Midjourney in median score but had a narrower IQR from 1.3 to 1.67. %

Analyzing scores by subject in \cref{fig:mcq-subject}, cat-themed images outperformed others with a median of 1.5 and an IQR from 1.33 to 1.84, while astronaut images lagged behind with a median of 1.33 and an IQR from 1.23 to 1.58. Next, dissecting the impact of an art background, \cref{fig:mcq-art-bg} shows that participants with art-related education or employment achieved slightly higher median scores at 1.5 and an IQR of 1.32 to 1.61 compared to those without, who had a median of 1.48 and a similar IQR but with a broader range and more outliers.
Participant demographics split by recruitment source, shown in \cref{fig:mcq-mturk}, highlight differences in scoring patterns. MTurk workers registered a median score of 1.48 and an IQR of 1.31 to 1.6, while university-recruited participants posted a slightly higher median of 1.54 with an IQR from 1.42 to 1.67. Notably, the MTurk group exhibited a broader score range and more outliers, indicating variances in scoring behavior between the two groups. 
These results broadly imply that when given with options, participants accurately identified subjects and modifiers in AI-generated images, achieving high scores across different models. %

\subsection{Human Inference Evaluation}
\label{sec:human-eval}

We next utilize metrics such as the image hash, perceptual similarity, CLIP score, semantic similarity, and survey similarity ratings (outlined in \cref{sec:metrics}) as applicable, to evaluate various factors affecting human inference accuracy. These metrics have ranges of 0 to 1 for CLIP score, semantic similarity, and perceptual similarity; and 0 to 64 for image hash score.

\subsubsection{Models.}
\label{sec:models-eval}

\begin{figure}[t]
\centering
\begin{subfigure}[b]{0.49\linewidth}
    \centering
    \includegraphics[width=\textwidth]{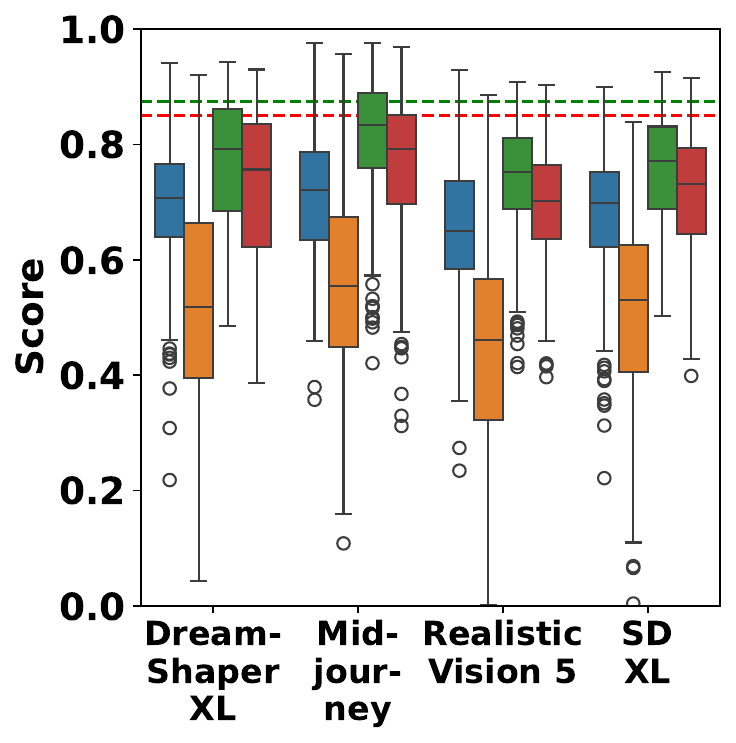}
    \caption{}
    \label{fig:model-controlled}
\end{subfigure}
\hfill
\begin{subfigure}[b]{0.49\linewidth}
    \centering
    \includegraphics[width=\textwidth]{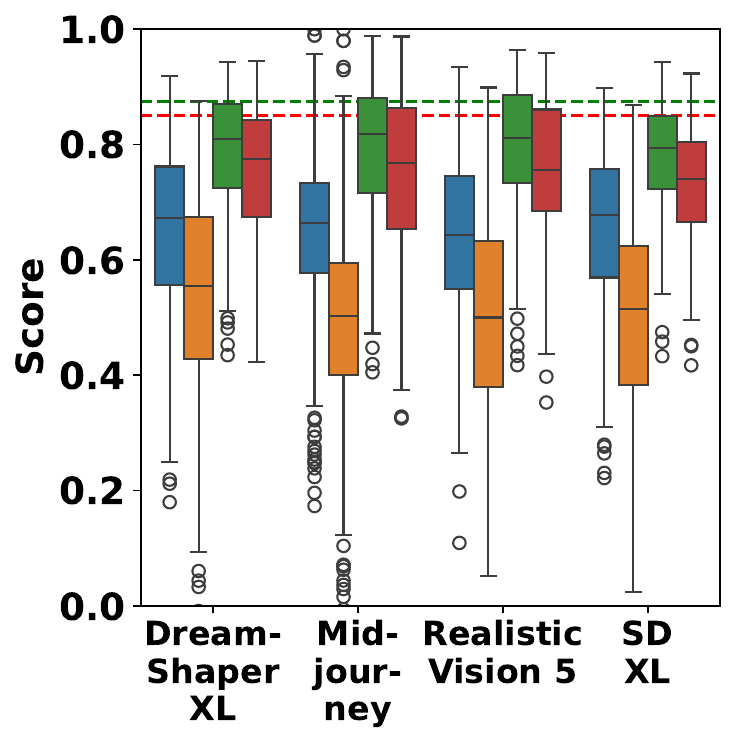}
    \caption{}
    \label{fig:model-uncontrolled}
\end{subfigure}
\hfill
\begin{subfigure}[b]{0.7\linewidth}
    \centering
    \includegraphics[width=\textwidth]{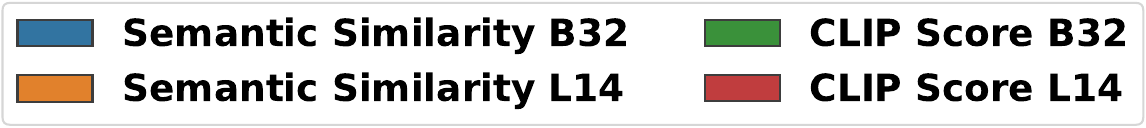}    
    \label{fig:model-legend}
\end{subfigure}
\caption{Semantic similarity and CLIP score for different \texttt{txt2img} models, in (a) controlled dataset, and (b) uncontrolled dataset. Success thresholds are depicted as dashed lines, green dashed line for the L14 model and red dashed line for the B32 model.}
\label{fig:model-clip}
\end{figure}

\begin{figure}[t]
\centering
\begin{subfigure}[b]{0.48\linewidth}
    \centering
    \includegraphics[width=\textwidth]{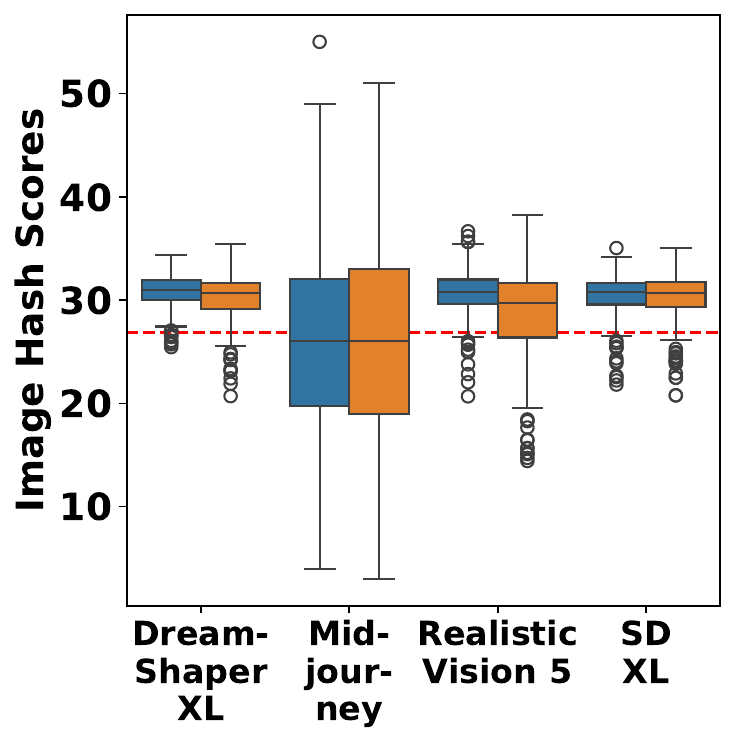}
    \label{fig:model-hash}
\end{subfigure}
\hfill
\begin{subfigure}[b]{0.48\linewidth}
    \centering
    \includegraphics[width=\textwidth]{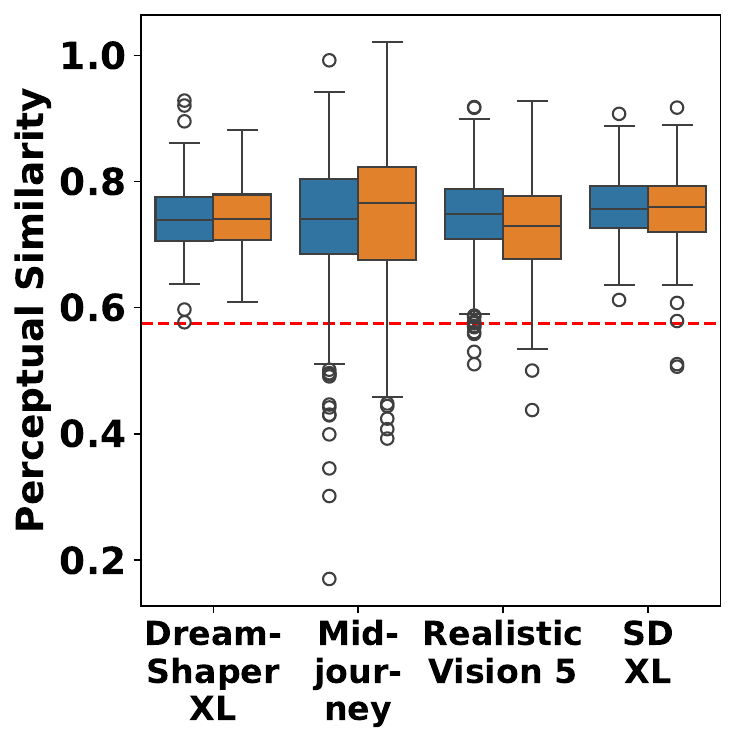}
    \label{fig:model-ps}
\end{subfigure}
\hfill
\begin{subfigure}[b]{0.6\linewidth}
    \centering
    \includegraphics[width=\textwidth]{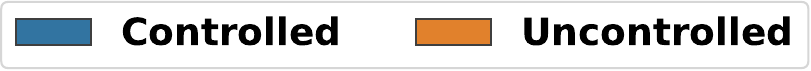}
    \label{fig:hash-ps-legend}
\end{subfigure}
\caption{Perceptual similarity and image hash scores for images generated using the four different \texttt{txt2img} models.}
\label{fig:model-hash-ps}
\end{figure}

\begin{figure}[t]
\centering
\begin{subfigure}[b]{0.99\linewidth}
    \centering
    \includegraphics[width=\textwidth]{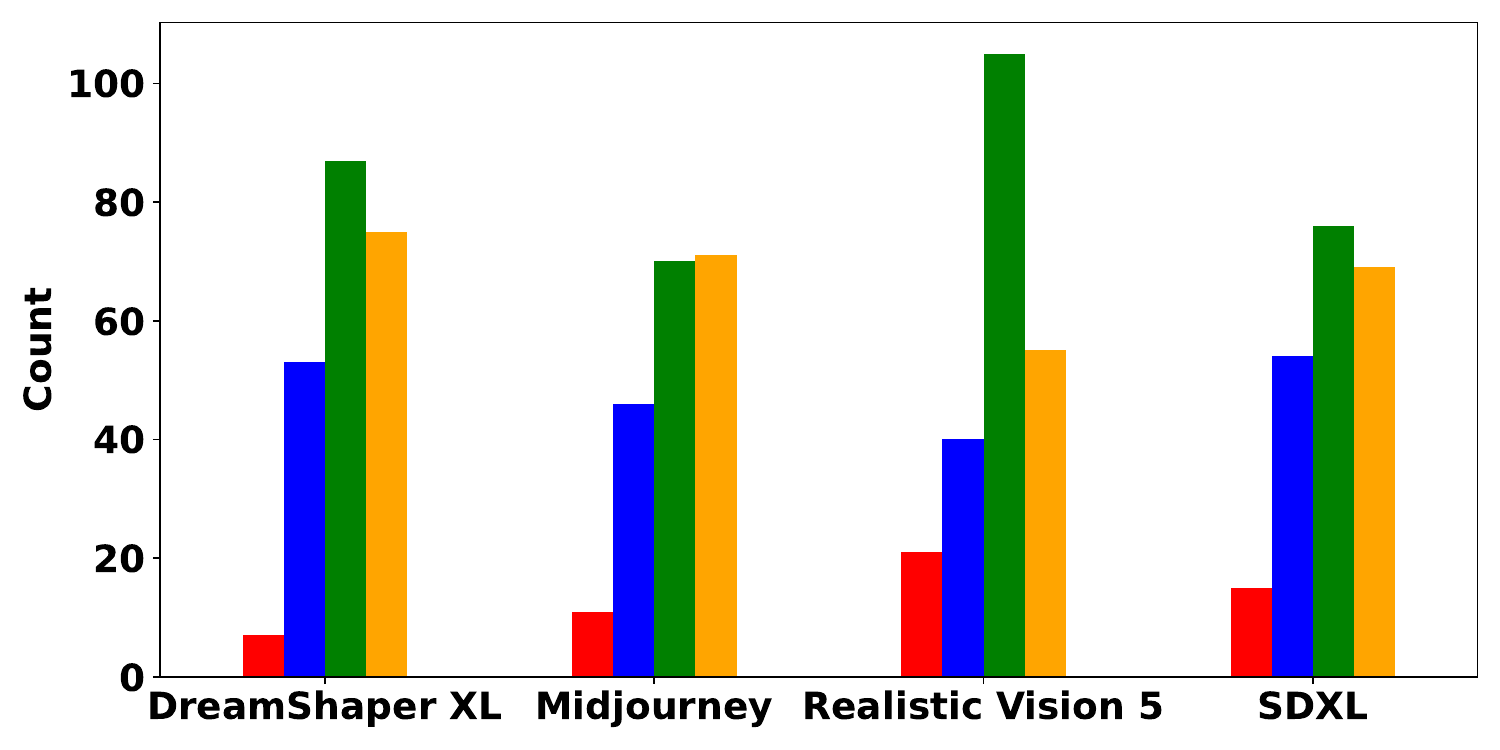}
    \label{fig:model-rating-count}
\end{subfigure}
\hfill
\begin{subfigure}[b]{0.7\linewidth}
    \centering
    \includegraphics[width=\textwidth]{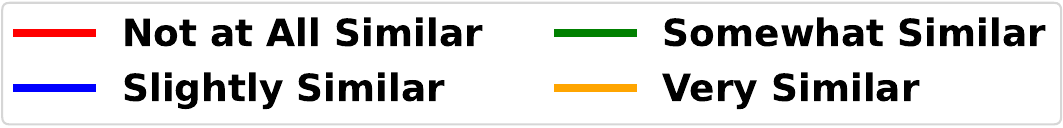}
    
    \label{fig:rating-legend}
\end{subfigure}
\caption{Part~IV similarity rating distributions for images generated using the four different \texttt{txt2img} models.} 
\label{fig:model-rating}
\end{figure}

Both the uncontrolled and controlled datasets (from Parts~II and III, respectively) serve as our primary data sources for this analysis where we compare the images generated from prompts submitted by the participants using our four selected \texttt{txt2img} models. %
These images are compared to the original image from the corresponding correct prompt using the the similarity metrics.  %

\Cref{fig:model-clip} shows the distribution of semantic similarity and CLIP scores obtained for the participant responses for survey Parts~II and III.
For the controlled dataset (\cref{fig:model-controlled}), semantic similarity (a higher value represents a higher similarity between the original and the participant inferred prompt) reveals that the L14 model falls short of the B32 model's performance, with a notable difference in median scores (25.687\% lower on average for L14).
This pattern persists in the CLIP scores (where a higher value represents a higher similarity between the original and the inferred prompt generated image), where B32 outperforms L14, emphasizing the latter's lower median by 5.294\% on average.
Among the \texttt{txt2img} models, Midjourney stands out for its superior semantic similarity and CLIP scores, indicating a closer match to the inferred prompts, whereas Realistic Vision 5 consistently records the lowest score ranges across these metrics.

In the uncontrolled dataset  (\cref{fig:model-uncontrolled}), we observe a similar trend, with the L14 model lagging behind the B32 model in both semantic similarity (the median is 22.06\% lower on average) and CLIP scores (the median 5.99\% lower on average)  (\cref{fig:model-uncontrolled}).
The majority of the CLIP scores observed fell below the success thresholds (dashed lines), $\theta_{\text{L14}} = 0.851$ and $\theta_{\text{B32}} = 0.875$, as denoted by the red and green dashed lines respectively in \cref{fig:model-clip}. However, it is noteworthy that \(162\) images from the Midjourney model out of \(2,286\) total images were able to surpass these thresholds under B32 across both controlled and uncontrolled datasets. %

The perceptual similarity and image hash scores, as shown in Figure \ref{fig:model-hash-ps}, reveal the influence of dataset conditions (controlled vs. uncontrolled) on image generation in a somewhat counter-intuitive manner.
Under controlled conditions, the hash scores are higher, with the median hash score across all models being on average 1.185\% higher than that for the uncontrolled dataset (note that a lower image hash scores indicate greater similarity).
Similarly, the median perceptual similarity is higher, although only marginally, at 0.329\% on average compared to the uncontrolled dataset (note that a lower perceptual similarity score is more desirable, i.e., greater similarity).
These findings indicate that a more structured approach to prompt inference results in the generation of images that considerably deviate from the original images compared to uncontrolled prompt inference. %

In terms of success thresholds, a pattern similar to CLIP scores was observed in the image hash and perceptual similarity evaluations. The success thresholds are depicted in \Cref{fig:model-hash} and \Cref{fig:model-ps} by red dashed lines. %
Surveyed similarity ratings (\cref{fig:model-rating}) provide direct insights into the perceived accuracy of images generated from human-inferred prompts. DreamShaper XL notably receives a majority of ``Somewhat Similar'' ratings (87), along with the highest count of ``Very Similar'' ratings at 75, suggesting that images generated with this model are likely to be more similar to the corresponding original image. 

These analyses reveals how different \texttt{txt2img} models perform while interpreting prompts and corresponding image generations, emphasizing the challenges in reproducing AI art through prompt inference that aligns with human interpretations.
Despite certain minor favorable observations, the overall performance across all three types of metrics did not reach the success thresholds, indicating the challenging nature in human prompt inferences. %

\subsubsection{Subjects.}
\label{sec:subjects-eval}

\begin{figure}[t]
\centering
\begin{subfigure}[b]{0.99\linewidth}
    \centering
    \includegraphics[width=\textwidth]{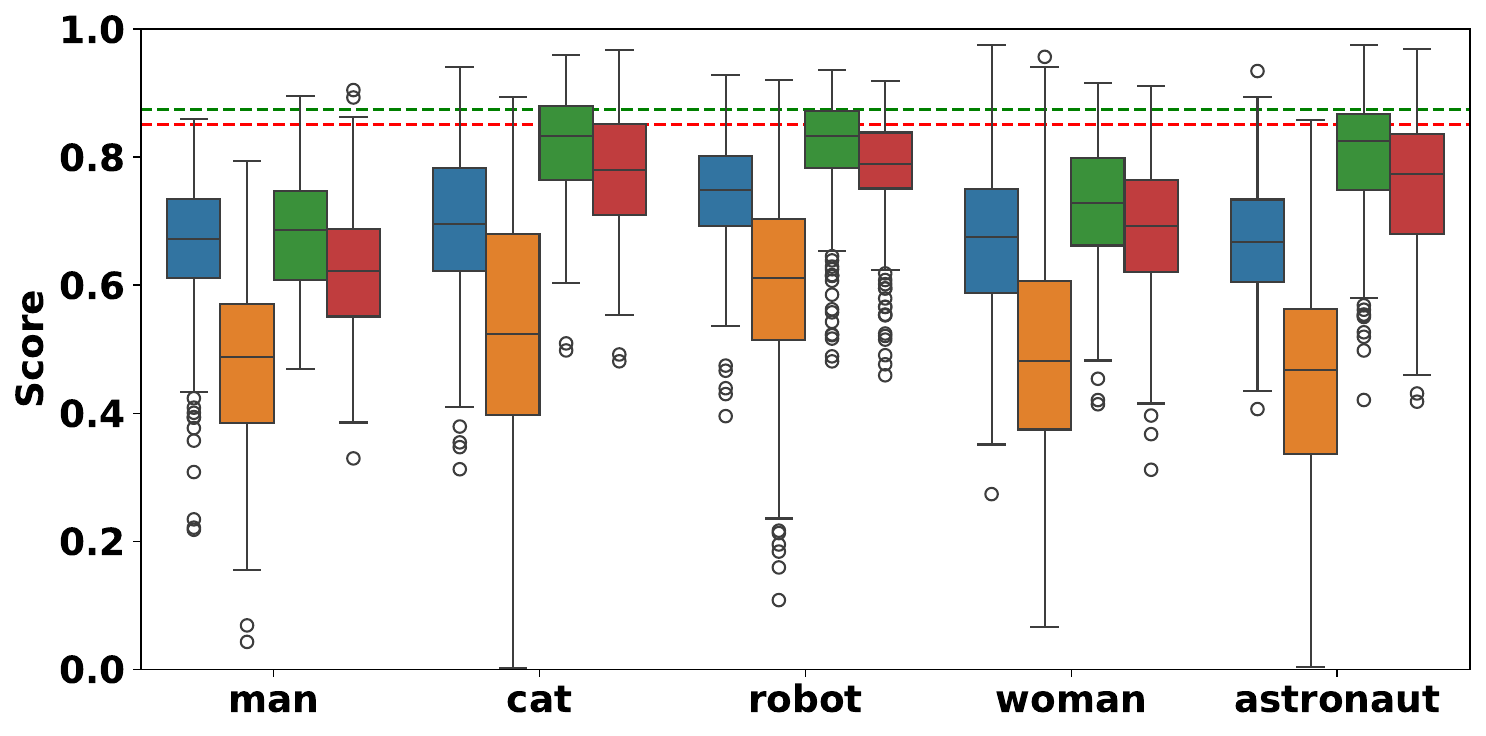}
    \label{fig:subject-controlled}
\end{subfigure}
\hfill
\begin{subfigure}[b]{0.7\linewidth}
    \centering
    \includegraphics[width=\textwidth]{figures/combined-graphs/by-models/clip-legend-crop.pdf}
    
    \label{fig:subject-legend}
\end{subfigure}
\caption{Semantic similarity and CLIP score for images containing different subjects, in controlled dataset.} 
\label{fig:subject-clip}
\end{figure}

\begin{figure}[t]
\centering
\begin{subfigure}[b]{0.48\linewidth}
    \centering
    \includegraphics[width=\textwidth]{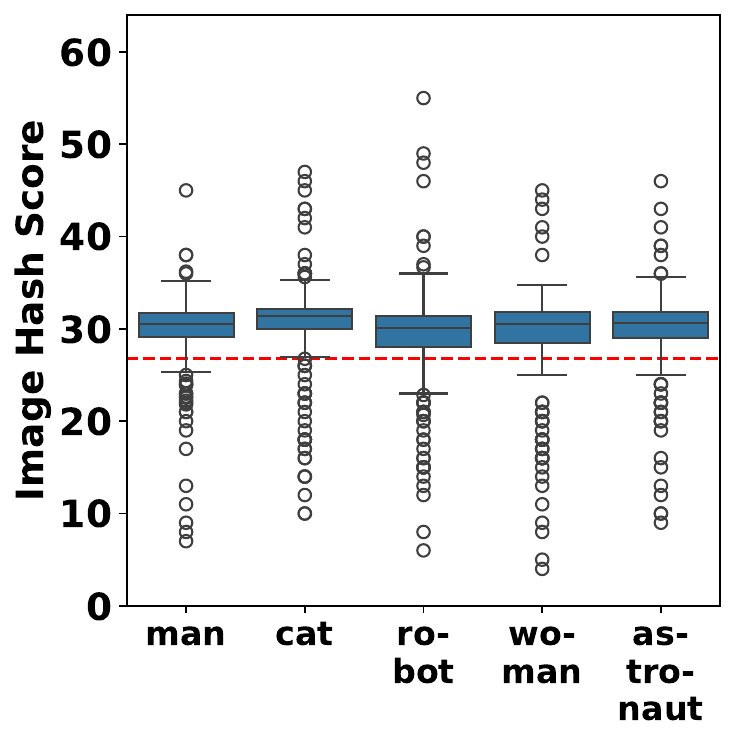}
    \label{fig:subject-hash}
\end{subfigure}
\hfill
\begin{subfigure}[b]{0.48\linewidth}
    \centering
    \includegraphics[width=\textwidth]{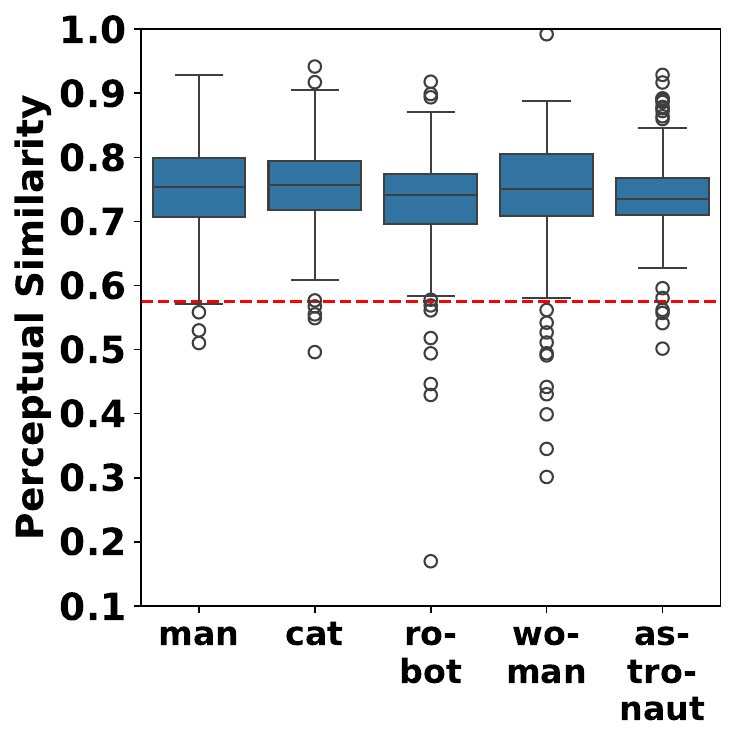}
    \label{fig:subject-ps}
\end{subfigure}
\caption{Perceptual similarity and image hash scores for images containing different subjects, in controlled dataset.}
\label{fig:subject-hash-ps}
\end{figure}

\begin{figure}[t]
\centering
\begin{subfigure}[b]{0.99\linewidth}
    \centering
    \includegraphics[width=\textwidth]{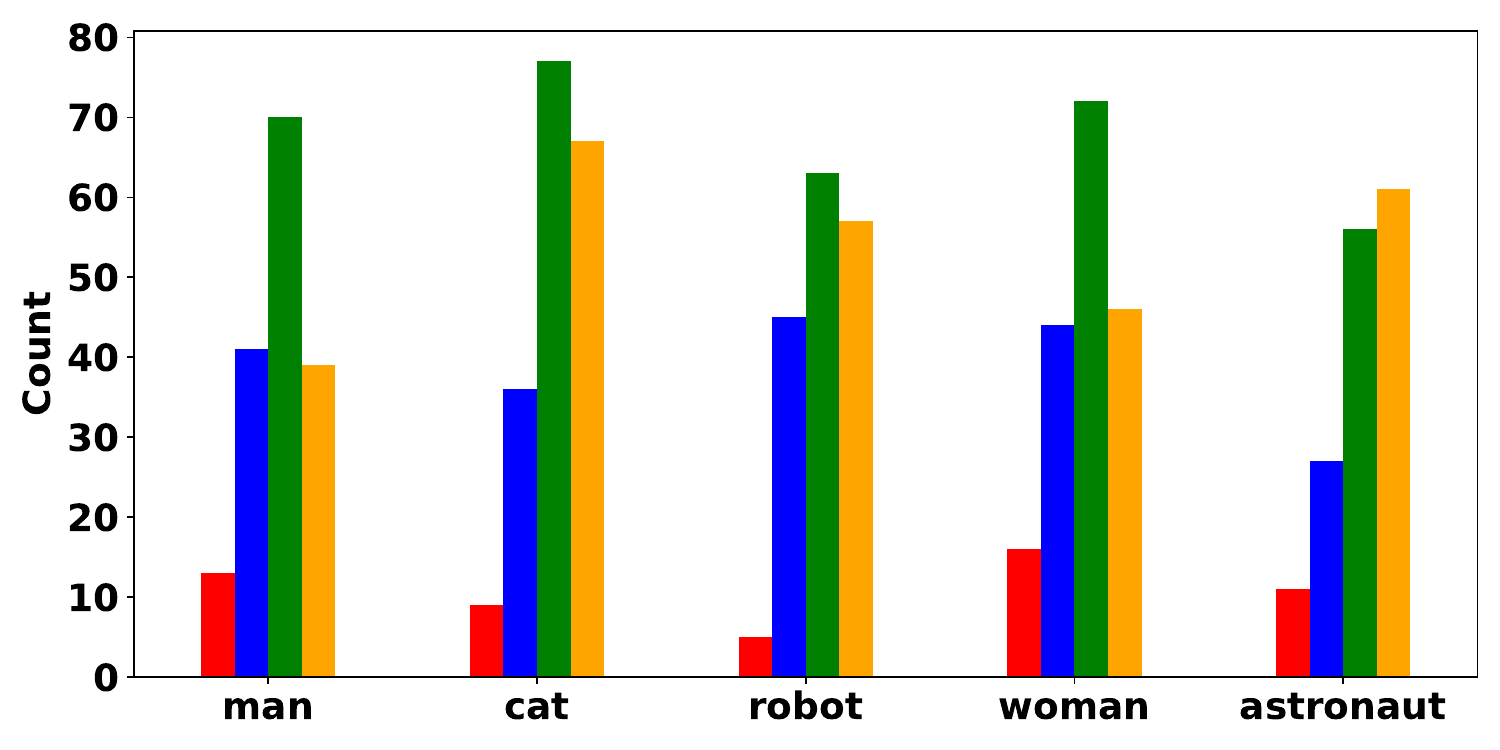}
    \label{fig:subject-rating-count}
\end{subfigure}
\hfill
\begin{subfigure}[b]{0.7\linewidth}
    \centering
    \includegraphics[width=\textwidth]{figures/part4_analysis/rating-legend-crop.pdf}
    
    \label{fig:rating-legend-subjects}
\end{subfigure}
\caption{Part~IV similarity rating distributions for images containing the five different subjects.} 
\label{fig:subject-rating}
\end{figure}

We next investigate the inference accuracy for different subjects in prompts using the controlled dataset. \Cref{fig:subject-clip} illustrates the semantic similarity and CLIP scores (where higher scores indicate a greater match), for five selected subjects. It is observed that the L14 model typically scores lower compared to the B32 model in both metrics. Subjects such as \textit{robots} and \textit{cats} achieve high CLIP scores, suggesting a strong alignment with their respective prompts. This observation prompts an inquiry into the nature of the subjects' representations: are they inherently distinct, consistently depicted by the models, or is it that humans are inherently better at inferring prompts related to these subjects? Inherent distinctness would imply that \textit{robots} and \textit{cats} possess unique, identifiable features that are readily recognized by the B32 and L14 models. Consistent depiction, on the other hand, would result in uniformity in representing these subjects across various image generations. Meanwhile, if humans are inherently better at inferring prompts for these particular subjects, this could also contribute to the observed accuracy.
While \textit{cats} may not always meet the desired benchmarks, in the top 25\% of cases, their inference performance reaches the success threshold, with 66 inference instances above $\theta_{\text{B32}}$ and 61 instances above $\theta_{\text{L14}}$. Except for \textit{man} and \textit{woman}, all subjects exhibit at least 30 instances surpassing the thresholds. %

Perceptual similarity and image hash scores, as depicted in \Cref{fig:subject-hash-ps}, indicate a consistent representation across subjects, although \textit{cats} show slightly higher scores, hinting at a less consistent inference process. 
For the image hash metric, only 30 to 40 instances for each subject show scores below the success threshold, indicating that lower scores, which signify better accuracy, are not frequently achieved.
However, the performance in terms of perceptual similarity is notably poorer across all subjects, with fewer than 12 instances falling below the success threshold. 

According to surveyed similarity ratings (\Cref{fig:subject-rating}), images of \textit{cats} and \textit{astronauts} predominantly received ``Very Similar'' ratings (67 and 61, respectively), denoting high inference accuracy. \textit{Robots} also achieved several ``Very Similar'' ratings (57) but with a noticeable amount of ``Slightly Similar'' ratings (45) as well, indicating some variability. Images depicting \textit{man} are mostly rated as ``Somewhat Similar,'' (70) which suggests moderate inference accuracy, with a distribution of ratings across different similarity levels, similar to images of \textit{woman}. %

Despite certain subjects such as \textit{cats} and \textit{robots} aligning moderately well in certain metrics, the overall analysis across subjects underscores that human prompt inference generally does not meet the success thresholds, emphasizing the inherent challenge of the task. These results suggest that while there are instances of alignment, the overall capability of humans to accurately infer prompts remains below par. %

\subsubsection{Modifiers.}
\label{sec:modifiers-eval}

\begin{figure}[t] %
\centering
\includegraphics[width=\linewidth]{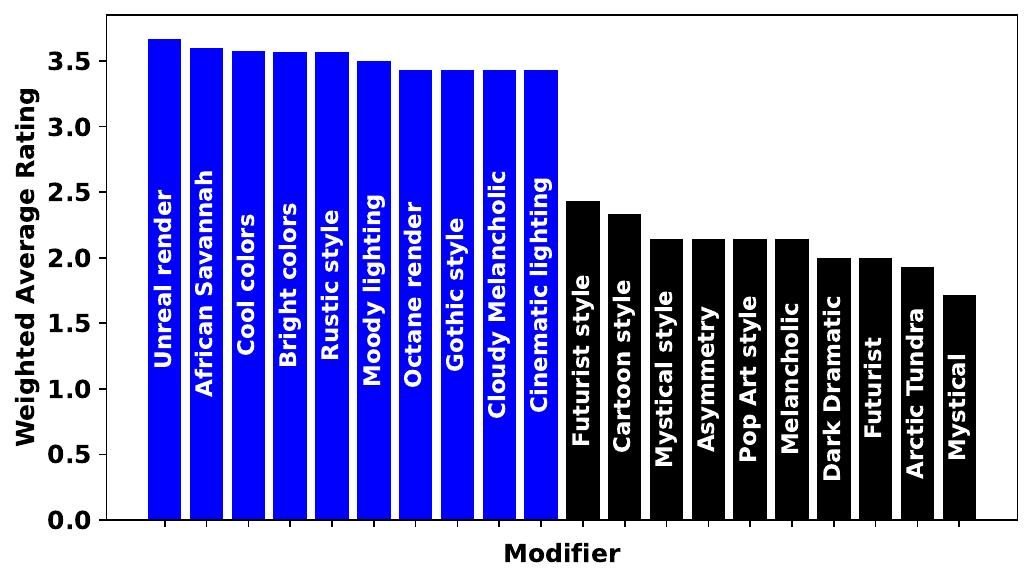} %
\caption{Top 10 and bottom 10 modifiers by weighted average ratings of the overall image similarity, by participants in Part~IV.}
\label{fig:top-bottom-mod}
\end{figure}

We next explored how modifiers present in the image generation prompts could affect the participants' prompt inference accuracy.
To better quantify this, we calculate a \textit{Weighted Average Rating} to assess the ranking of the modifiers through the images shown in Part IV. %
This metric is determined by assigning a normalized value to each familiarity rating, where \(1\) represents ``Not at All Similar'' and \(4\) represents ``Very Similar''. The process involves counting the frequency of each rating, multiplying it by its normalized value, and then calculating the average of these products. Formally, the Weighted Average Rating, \(W\), is expressed as:
\[
W = \frac{\sum_{i=1}^{n} f_i \cdot v_i}{\sum_{i=1}^{n} f_i}
\]

where \(f_i\) is the frequency of the \(i\)-th rating, \(v_i\) is the normalized value of the \(i\)-th rating, and \(n\) is the number of rating types. %
\cref{fig:top-bottom-mod} shows 20 modifiers from our list of 121, the top 10 modifiers with the highest weighted average, in blue; and the bottom 10 modifiers, with the lowest weighted average, in black.%
Notably, some of the top modifiers such as ``Unreal render,'' and ``Cloudy Melancholic'' are potentially challenging to be inferred directly by participants. This observation suggests two possibilities: either the top modifiers exert substantial influence, making them easily identifiable and thus straightforward to infer by participants, or, despite their complexity, these modifiers do not significantly impact the image generation process, thereby not detrimentally affecting the participants' ability to recognize the intended imagery. %

\subsubsection{Art Background vs. Other Backgrounds.}
\cref{fig:art-bg-clip} in \cref{appendix-additional-results} shows that participants with an art background tend to introduce more variability and slightly lower scores in semantic similarity and CLIP scores across both controlled (where the median of semantic similarity from B32 is 0.465\% lower on average and median CLIP score of B32 is 0.172\% lower on average) and uncontrolled (where the median of semantic similarity from B32 is 0.292\% lower on average, CLIP score B32 is 0.057\% lower on average) datasets. This suggests that their refined ability to discern certain image characteristics may lead to more critical or complex interpretations of the AI-generated images, potentially diverging from the target images more than those without an art background.

Participants for both groups have their CLIP score below the success thresholds (green and red dashed lines in \cref{fig:art-bg-clip} indicate the thresholds $\theta_{\text{B32}}$ and $\theta_{\text{L14}}$ respectively.). 
In the controlled data set, only 1 out of 134 participants in the Non-Art were able to have an average CLIP B32 score above the threshold; while the inverse is true for the uncontrolled dataset; where, notably, there were only 1 participant with Art background (out of 36) and 13 Non-Art participants (out of 134) were able to achieved scores above the thresholds. 

Perceptual similarity and image hash scores (\cref{fig:art-bg-hash-ps}) show negligible differences between the art and non-art groups (e.g. perceptual similarity median only shows a 0.02\% difference between the group in the uncontrolled dataset, and 0.326\% difference in the controlled dataset).
Surveyed similarity ratings (\cref{fig:art-bg-rating}) also does not provide any significant difference between two participant groups. 
In summary, the involvement of individuals with certain expertise in the field, such as those with an art background, does not appear to significantly enhance the accuracy of human prompt inference. %

\subsubsection{MTurk Workers vs. University Population.}
\cref{fig:mturk-clip} in \cref{appendix-additional-results} indicates that university participants generally achieved higher median scores across all categories compared to MTurk workers (semantic similarity B32 is 3.168\% higher and CLIP score B32 is 4.311\% higher for uncontrolled dataset on average, while the controlled dataset sees the semantic similarity B32 to be 2.2\% higher, and CLIP score B32 to be be 4.446\% higher on average), suggesting a more accurate inference of prompts. This difference is particularly notable in the controlled dataset, where the tasks may be inherently more straightforward due to the structured nature of the prompts. However, a significant observation is the broader range of scores among MTurk participants, pointing to a greater variability in their prompt interpretations.
In terms of success thresholds, participants for both groups have their CLIP scores below the success threshold. In the uncontrolled dataset, there were 7 participants from either group who achieved a CLIP B32 scores above the success threshold. While in the controlled dataset, there were only 1 out of 59 participants in the university group with their score above the threshold.   

Perceptual similarity and image hash scores (\cref{fig:mturk-hash-ps} in \cref{appendix-additional-results}) further supports this observation, showing that median scores are closely aligned between university and MTurk participants. %
In summary, while university participants tend to perform better on average, indicating potentially more precise prompt inferences, the variability among MTurk workers highlights a diverse range of interpretations and evaluations.

\subsubsection{AI Tool Familiarity.} %
\label{sec:tool-familiarity-score}

Participants with limited familiarity with generative AI tools exhibited higher inference accuracy across metrics, such as higher B32 and L14 CLIP scores (\cref{fig:controlled-tool-b32-score,fig:controlled-tool-l14-score} in \cref{appendix-additional-results}), as well as lower image hash and perceptual similarity scores (\cref{fig:controlled-hash-score,fig:controlled-tool-ps-score}) in the controlled dataset. This outcome may suggest that participants without extensive AI tool exposure are either applying broader interpretations that align well with AI-generated images or that their fresh intuition-based prompt inference aligns better with the \texttt{txt2img} models' generative behavior.
The uncontrolled dataset reaffirms this trend, where participants with minimal AI tool familiarity again demonstrated slightly higher prompt inference accuracy (\cref{fig:uncontrolled-tool-b32-score,fig:uncontrolled-tool-l14-score,fig:uncontrolled-hash-score,fig:uncontrolled-tool-ps-score}). 
Being more familiar with generative AI tools does not seem to affect how close the participant's score is to the success thresholds and overall all familiarity categories still fall short of the success thresholds across all the metrics.

The cumulative findings thus far indicate that human performance in prompt inference is generally below par, regardless of the image generation models, subjects or modifiers involved, or the demographic backgrounds of the participants. Yet, before drawing any definitive conclusions, we will explore the potential for enhancing prompt inference accuracy through a synergistic approach that combines AI and human capabilities.

\subsection{Human-AI Combined Inference}
\label{sec:colab-eval}

\begin{figure}[t]
\centering
\begin{subfigure}[b]{0.49\linewidth}
    \centering
    \includegraphics[width=\textwidth]{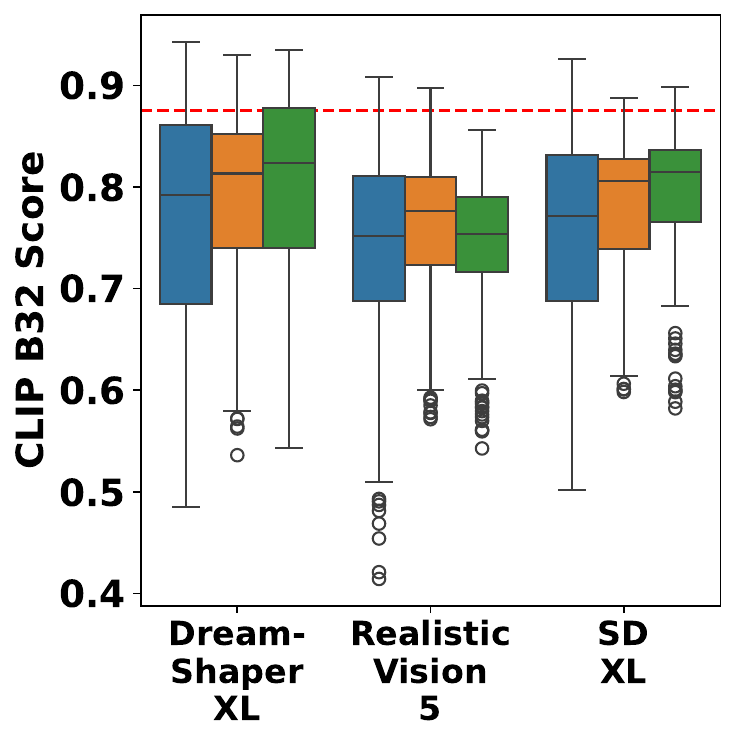}
    \caption{}
    \label{fig:collab-b32}
\end{subfigure}
\hfill
\begin{subfigure}[b]{0.49\linewidth}
    \centering
    \includegraphics[width=\textwidth]{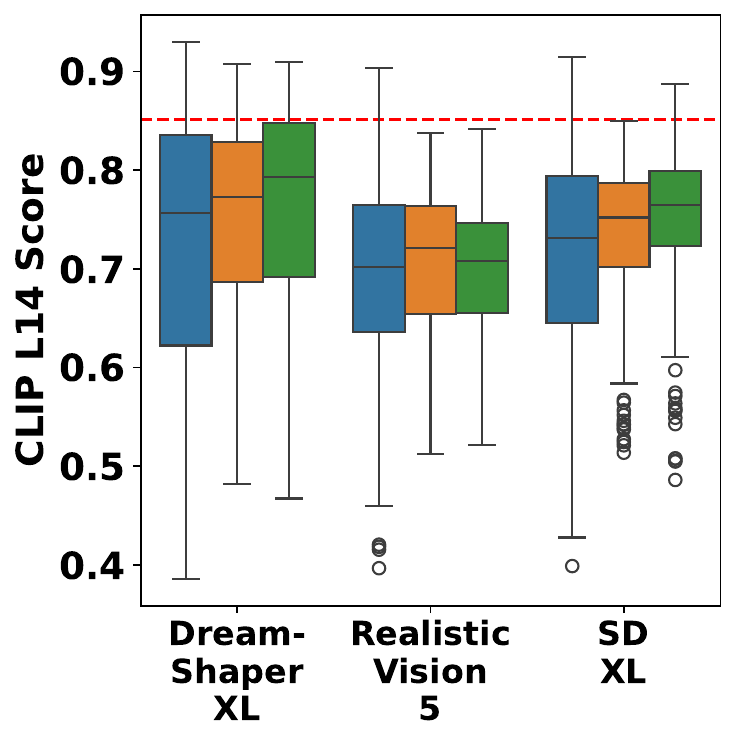}
    \caption{}
    \label{fig:collab-l14}
\end{subfigure}
\hfill
\begin{subfigure}[b]{0.8\linewidth}
    \centering
    \includegraphics[width=\textwidth]{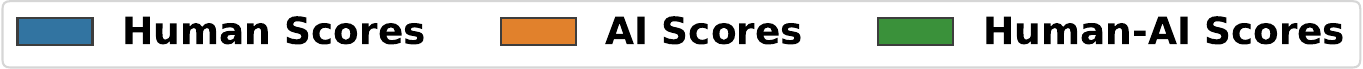}
    \label{fig:collab-clip-legend}
\end{subfigure}
\caption{B32 (a) and L14 (b) CLIP scores for different models from prompts generated by human, AI, and human-AI combination.}
\label{fig:collab-clip}
\end{figure}

\begin{figure}[t]
\centering
\begin{subfigure}[b]{0.49\linewidth}
    \centering
    \includegraphics[width=\textwidth]{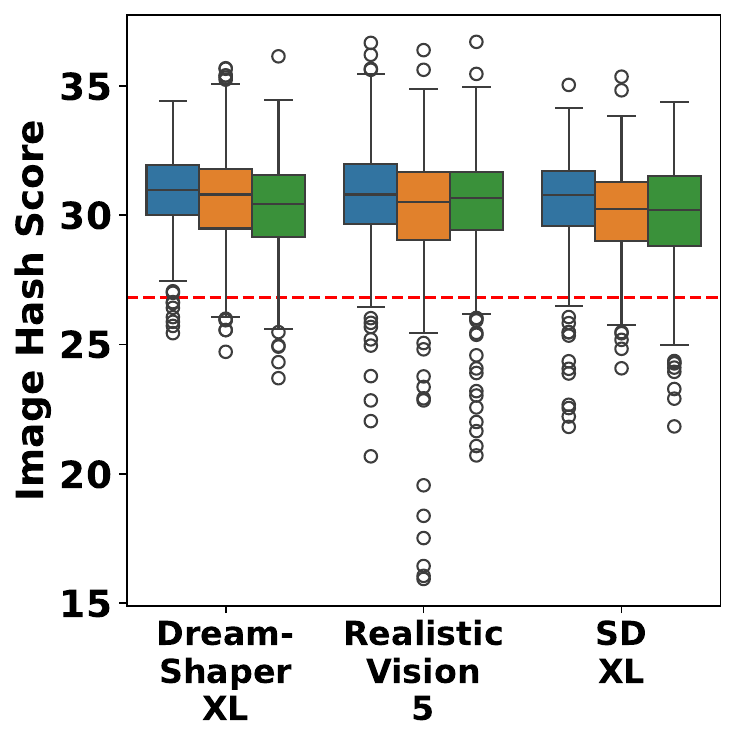}
    \caption{}
    \label{fig:collab-hash}
\end{subfigure}
\hfill
\begin{subfigure}[b]{0.49\linewidth}
    \centering
    \includegraphics[width=\textwidth]{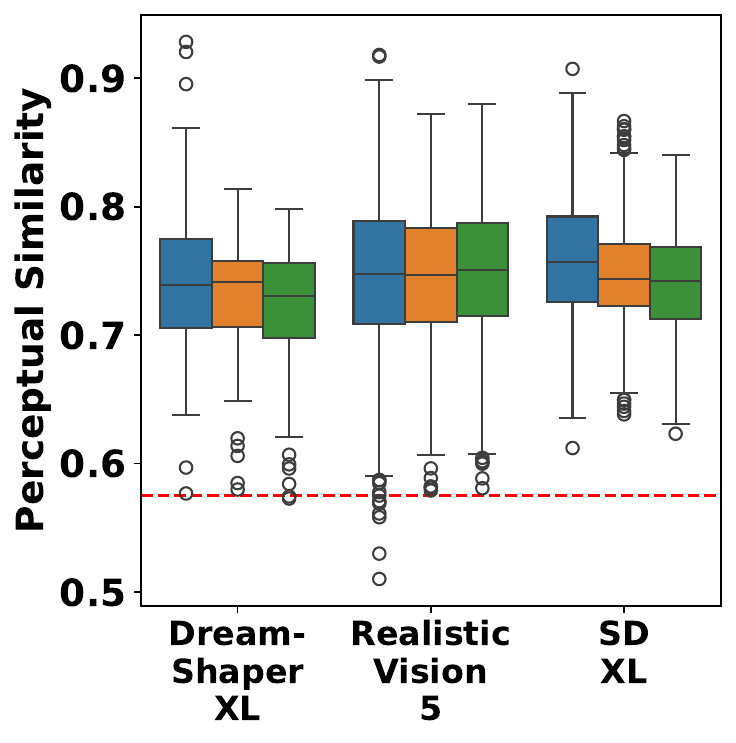}
    \caption{}
    \label{fig:collab-ps}
\end{subfigure}
\hfill
\begin{subfigure}[b]{0.8\linewidth}
    \centering
    \includegraphics[width=\textwidth]{figures/human-ai-collab_analysis/image_similarity/human-ai-legend-crop.pdf}
    \label{fig:collab-legend}
\end{subfigure}
\caption{Image hash score (a) and perceptual similarity (b) for different models from prompts generated by human, AI, and human-AI combination.}
\label{fig:collab-hash-ps}
\end{figure}

\begin{figure}[t]
\centering
\begin{subfigure}[b]{0.49\linewidth}
    \centering
    \includegraphics[width=\textwidth]{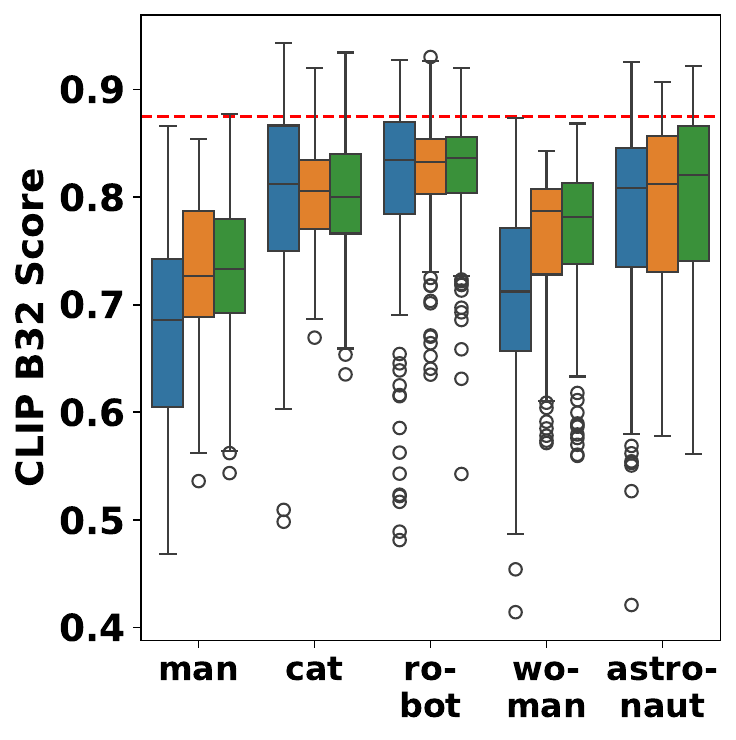}
    \caption{}
    \label{fig:collab-subject-b32}
\end{subfigure}
\hfill
\begin{subfigure}[b]{0.49\linewidth}
    \centering
    \includegraphics[width=\textwidth]{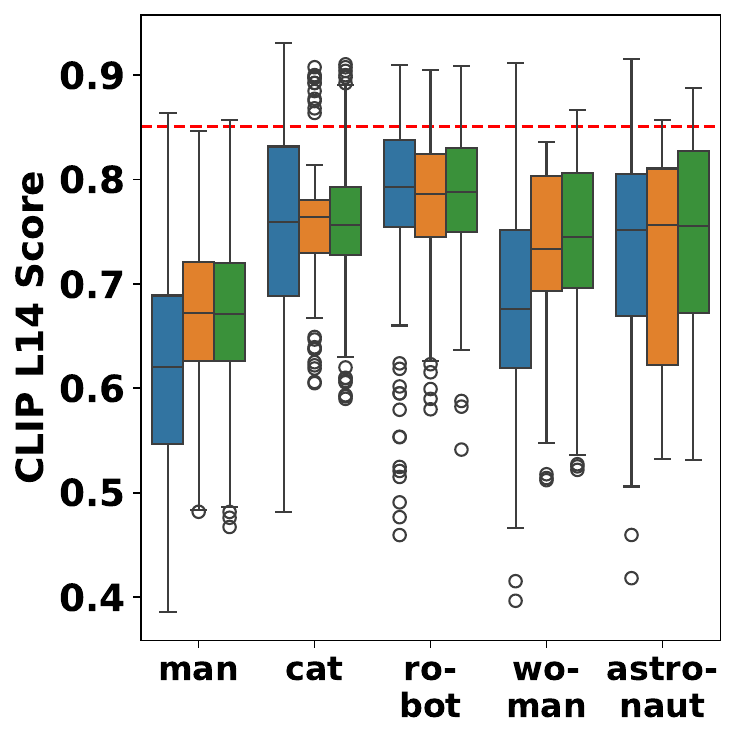}
    \caption{}
    \label{fig:collab-subject-l14}
\end{subfigure}
\hfill
\begin{subfigure}[b]{0.8\linewidth}
    \centering
    \includegraphics[width=\textwidth]{figures/human-ai-collab_analysis/image_similarity/human-ai-legend-crop.pdf}
    \label{fig:collab-subject-legend}
\end{subfigure}
\caption{B32 (a) and L14 (b) CLIP scores for different subjects and modifiers from prompts generated by human, AI, and human-AI combination.}
\label{fig:collab-subject-clip}
\end{figure}

\begin{figure}[t]
\centering
\begin{subfigure}[b]{0.49\linewidth}
    \centering
    \includegraphics[width=\textwidth]{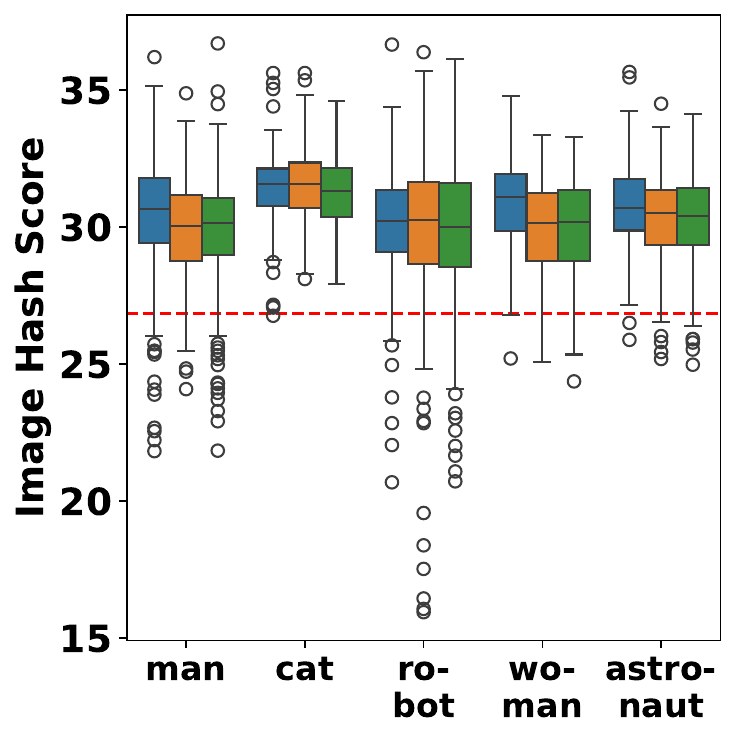}
    \caption{}
    \label{fig:collab-hash-subject}
\end{subfigure}
\hfill
\begin{subfigure}[b]{0.49\linewidth}
    \centering
    \includegraphics[width=\textwidth]{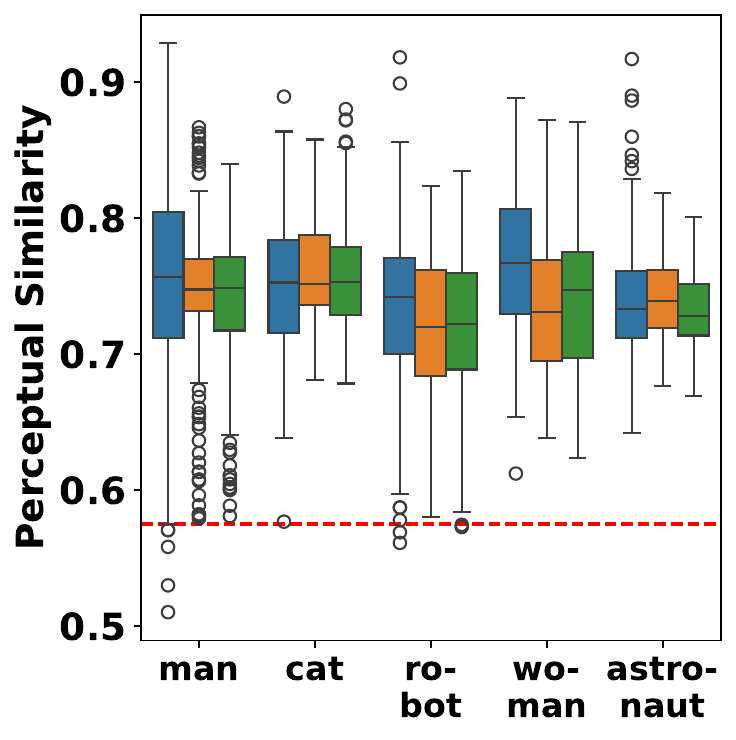}
    \caption{}
    \label{fig:collab-ps-subject}
\end{subfigure}
\hfill
\begin{subfigure}[b]{0.8\linewidth}
    \centering
    \includegraphics[width=\textwidth]{figures/human-ai-collab_analysis/image_similarity/human-ai-legend-crop.pdf}
    \label{fig:collab-hash-ps-subjectlegend}
\end{subfigure}
\caption{Image hash score (a) and perceptual similarity score (b) for different subjects and modifiers from prompts generated by human, AI, and human-AI combination.}
\label{fig:collab-hash-ps-subject}
\end{figure}

We next examine the effectiveness of combining human inferences with AI inferences, particularly using CLIP interrogator for inference and then merging them using GPT-4 (as outlined in \cref{sec:human-ai-expsetup}), towards accurately inferring prompts for AI-generated images. 
CLIP score analysis shows that combination of humans and AI inferred prompts generally results in higher image CLIP scores (\cref{fig:collab-clip}).
Specifically, the B32 scores were found to be 3.126\% higher  
on average in a combined setting compared to individual efforts.  %
However, when compared to the images generated by prompts inferred by CLIP Interrogator, the B32 score saw a 0.23\% decrease on average, due to the Realistic Vision 5-generated images performing worse; although Midjourney-generated images performed 1.265\% better, and SDXL images with a 1.099\% increase in performance (\Cref{fig:collab-b32}).
With that being said, the overall improvement highlights how combining human inference with AI inference can lead to slightly better accuracy in matching prompts with generated images. The overlap in scores across different \texttt{txt2img} models suggests that while human-AI combined prompts enhances accuracy, the degree of improvement may not be drastically distinct among different AI models.

Perceptual similarity and image hash score (\cref{fig:collab-hash-ps}) also indicate that images generated through human-AI combined prompts align more closely with target images, as evident from the slightly lower average hash scores (by 1.417\%) and perceptual similarity scores (by 0.905\%). \Cref{fig:collab-subject-clip,fig:collab-hash-ps-subject} further explores this combined prompt inference, showcasing that human-AI combined prompts not only consistently improves inference across various subjects but also indicates a nuanced enhancement in how specific subjects and modifiers are interpreted and matched (hash score median is on average 1.438\% lower, while perceptual similarity median is on average 1.431\% lower).

For all 4 types of metrics, combining AI inferred prompts through CLIP interrogator with human inferred prompt has a positive effect on the overall accuracy. While the human-AI combined prompts still did not reach the respective metrics' success thresholds, its performance is slightly better than that of purely human inferred prompts. This suggests that accurate prompt inference by a human would benefit from combination with AI inference.

The enhanced accuracy in prompt inference through the combination of human and AI efforts %
can be attributed to the complementary skills and knowledge each brings. Humans excel in creative and contextual understanding, while AI provides extensive data analysis and pattern recognition capabilities. This combination allows for error reduction, iterative refinement, and a broader knowledge base, leading to more accurate prompts. Despite these benefits, the fact that performance did not reach the success thresholds suggests that further optimizations in human-AI interaction are required.%

In summary, the empirical evidence gathered from the evaluation, shows that while humans alone may struggle with prompt inference, their efforts, when combined with AI, demonstrate a modest improvement. The findings suggest there's room for growth in human-AI collaborations, which could eventually enhance the efficacy of prompt reconstruction. Until such advancements are realized, prompt marketplaces maintain their practicality as a business model in the realm of AI-generated art.%

\section*{Conclusion}
\label{sec:conclusion}

The study presented in this paper explores the intellectual property considerations of prompts in the realm of AI-generated art, particularly focusing on the ability of humans to infer these prompts by solely examining the resulting artworks.
The below par results from our human only prompt inference experiments suggest that the performance of human participants is influenced by the complexities in the prompt and the generated artwork. However, the combination of prompts from AI models such as CLIP interrogator has a positive effect on improving the inference efforts of the human participants.
Collectively, these observations point towards the feasibility of generative AI prompt marketplaces as viable business models, where the uniqueness and creativity of prompts can be valued and traded while maintaining their secure intellectual property.
Moreover, the exploration of combining human insights with AI in improving prompt inference opens new avenues for research. %

\bibliography{references}

\subsection*{Paper Checklist}
\label{sec:checklist}

\begin{enumerate}
\item For most authors...
\begin{enumerate}
    \item  Would answering this research question advance science without violating social contracts, such as violating privacy norms, perpetuating unfair profiling, exacerbating the socio-economic divide, or implying disrespect to societies or cultures?
    \answerYes{Yes, this survey should not have any of the negative impact mentioned above.}
  \item Do your main claims in the abstract and introduction accurately reflect the paper's contributions and scope?
    \answerYes{Yes}
   \item Do you clarify how the proposed methodological approach is appropriate for the claims made? 
    \answerNA{NA}
   \item Do you clarify what are possible artifacts in the data used, given population-specific distributions?
    \answerNA{NA}
  \item Did you describe the limitations of your work?
    \answerNA{NA}
  \item Did you discuss any potential negative societal impacts of your work?
    \answerNo{No}
      \item Did you discuss any potential misuse of your work?
    \answerNA{NA}
    \item Did you describe steps taken to prevent or mitigate potential negative outcomes of the research, such as data and model documentation, data anonymization, responsible release, access control, and the reproducibility of findings?
    \answerNA{NA}
  \item Have you read the ethics review guidelines and ensured that your paper conforms to them?
    \answerYes{Yes}
\end{enumerate}

\item Additionally, if your study involves hypotheses testing...
\begin{enumerate}
  \item Did you clearly state the assumptions underlying all theoretical results?
    \answerNA{NA}
  \item Have you provided justifications for all theoretical results?
    \answerNA{NA}
  \item Did you discuss competing hypotheses or theories that might challenge or complement your theoretical results?
    \answerNA{NA}
  \item Have you considered alternative mechanisms or explanations that might account for the same outcomes observed in your study?
    \answerNA{NA}
  \item Did you address potential biases or limitations in your theoretical framework?
    \answerNA{NA}
  \item Have you related your theoretical results to the existing literature in social science?
    \answerNA{NA}
  \item Did you discuss the implications of your theoretical results for policy, practice, or further research in the social science domain?
    \answerNA{NA}
\end{enumerate}

\item Additionally, if you are including theoretical proofs...
\begin{enumerate}
  \item Did you state the full set of assumptions of all theoretical results?
    \answerNA{NA}
	\item Did you include complete proofs of all theoretical results?
    \answerNA{NA}
\end{enumerate}

\item Additionally, if you ran machine learning experiments...
\begin{enumerate}
  \item Did you include the code, data, and instructions needed to reproduce the main experimental results (either in the supplemental material or as a URL)?
    \answerNA{NA}
  \item Did you specify all the training details (e.g., data splits, hyperparameters, how they were chosen)?
    \answerNA{NA}
     \item Did you report error bars (e.g., with respect to the random seed after running experiments multiple times)?
    \answerNA{NA}
	\item Did you include the total amount of compute and the type of resources used (e.g., type of GPUs, internal cluster, or cloud provider)?
    \answerNA{NA}
     \item Do you justify how the proposed evaluation is sufficient and appropriate to the claims made? 
    \answerNA{NA}
     \item Do you discuss what is ``the cost`` of misclassification and fault (in)tolerance?
    \answerNA{NA}
  
\end{enumerate}

\item Additionally, if you are using existing assets (e.g., code, data, models) or curating/releasing new assets, \textbf{without compromising anonymity}...
\begin{enumerate}
  \item If your work uses existing assets, did you cite the creators?
    \answerNA{NA}
  \item Did you mention the license of the assets?
    \answerNA{NA}
  \item Did you include any new assets in the supplemental material or as a URL?
    \answerNA{NA}
  \item Did you discuss whether and how consent was obtained from people whose data you're using/curating?
    \answerNA{NA}
  \item Did you discuss whether the data you are using/curating contains personally identifiable information or offensive content?
    \answerNA{NA}
\item If you are curating or releasing new datasets, did you discuss how you intend to make your datasets FAIR?
\answerNA{NA}
\item If you are curating or releasing new datasets, did you create a Datasheet for the Dataset? 
\answerNA{NA}
\end{enumerate}

\item Additionally, if you used crowdsourcing or conducted research with human subjects, \textbf{without compromising anonymity}...
\begin{enumerate}
  \item Did you include the full text of instructions given to participants and screenshots?
    \answerYes{Yes, details the Appendix.}
  \item Did you describe any potential participant risks, with mentions of Institutional Review Board (IRB) approvals?
    \answerNo{No, the study was granted an Exempt status, and posed no risks to participants.}
  \item Did you include the estimated hourly wage paid to participants and the total amount spent on participant compensation?
    \answerNA{NA}
   \item Did you discuss how data is stored, shared, and deidentified?
   \answerNA{NA}
\end{enumerate}

\end{enumerate}

\appendix
\section{Prompts and Marketplaces}
\label{appendix:prompts-marketplaces}

\subsection{Prompt Marketplaces}

\begin{figure}[H]
\centering
\includegraphics[width=0.99\linewidth]{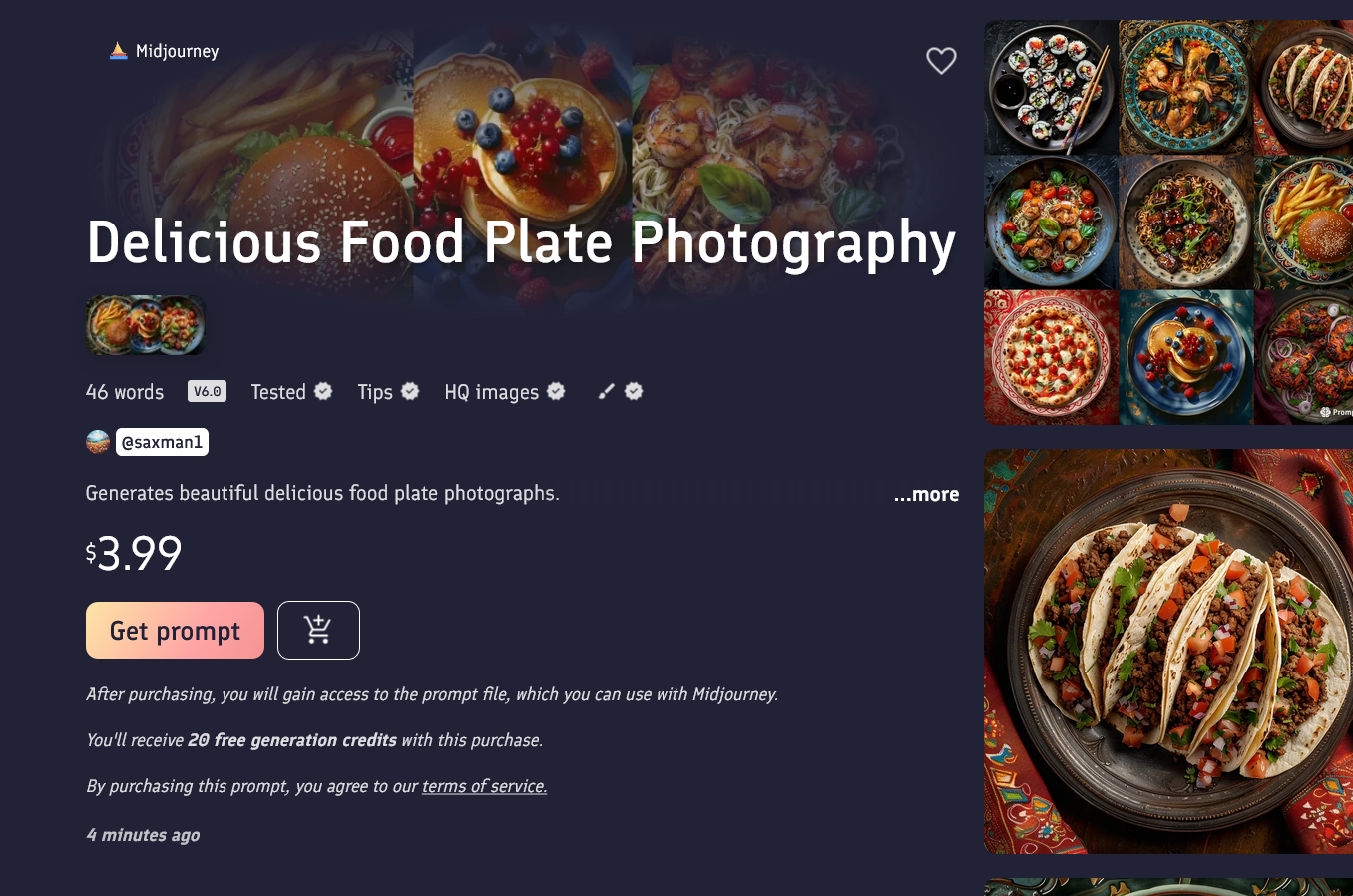}
\caption{A PromptBase listing selling the prompt to a specific style of artwork.}%
\label{fig:promptbase-example}
\end{figure}

The integration of \texttt{txt2img} models in digital art creation has led to the development of AI prompt marketplaces, such as PromptHero\footnote{\url{https://prompthero.com}}, Promptrr.io\footnote{\url{https://promptrr.io}}, Prompti AI\footnote{\url{https://prompti.ai}}, PromptBase\footnote{\url{https://promptbase.com}}, and 
CivitAI\footnote{\url{https://civitai.com}}. These platforms facilitate the buying, selling, or sharing of prompts designed for various \texttt{txt2img} models (\cref{fig:promptbase-example}).
Prompt marketplaces operate on a business model in which users can create and submit their own prompts or purchase those created by others. This has turned prompt creation into a profitable activity, as effective prompts greatly enhance the quality of AI-generated art~\cite{decrypt.coLucrative}. Users selling prompts on these platforms can earn income, while buyers gain access to a diverse range of ready-to-use prompts, enhancing their productivity and creativity in AI art generation.
These marketplaces typically use a commission-based revenue model, where prompt authors pay the platform a part of their revenue whenever the platform sells their prompt to a customer~\cite{makeuseofAIprompts}. These platforms also generally treat prompts as intellectual property~\cite{promptbasetos,promptrrtos}, recognizing and protecting the creative effort involved in their creation.

\subsection{Subjects and Modifiers}
\label{sec:back-subjects-modifiers}

The use of \textit{subjects} and \textit{modifiers} in prompts plays an important role in guiding a \texttt{txt2img} model to produce images that align with the creator's vision. %
The subject serves as the core theme or focus of the image, while modifiers provide the specifications that refine and shape the final output. 
Subjects can range from concrete objects, like ``cat,'' ``forest'' or ``robot,'' to more abstract concepts, such as ``solitude'' or ``chaos.'' On the other hand, the modifiers adjust the images's aesthetic by specifying attributes such as color, texture, time of day, or emotional ambiance, effectively guiding the AI towards a more targeted and refined artistic rendition. \cref{fig:sub-mod-example} demonstrates an example of how adding different modifiers alongside a subject (cat) produces varied image generations.
Prompts for \texttt{txt2img} models are typically limited by 300 to 400 input characters (depending on the model) on how many subjects and modifiers can be specified for each image generation.

\section{Analyzed \texttt{txt2img} Model Details}
\label{appendix:txt2img-models-details}

\begin{itemize}[leftmargin=*]
\item \textbf{MidJourney v5.0}~\cite{midjourney} from Midjourney Inc. is a closed-source \texttt{txt2img} model which is recognized for generating images that often exhibit a unique artistic and abstract quality. Due to its closed-source nature, not much is known about the architecture and parameters of its image generator model or the dataset used to train it. MidJourney v5.0 can natively generate images of 1024$\times$1024 pixels, using a Discord bot with \texttt{/imagine [Prompt]} command. %

\item \textbf{Stable Diffusion XL} (SDXL)~\cite{podell2023sdxl,rombach2022high}, created by Stability AI, employs a variational autoencoder (VAE) in conjunction with a cross-attention transformer-based architecture as the generator. %
Stable Diffusion XL can also natively generate images of 1024$\times$1024 pixels, using direct interface with the publicly available model~\cite{sdxlhug}. %

\item \textbf{DreamShaper XL} was trained and fine-tuned over Stable Diffusion XL~\cite{podell2023sdxl}, with a focus on generating photo-realistic fantasy images. Similar to Stable Diffusion XL, DreamShaper XL can also natively generate images of 1024$\times$1024 pixels.

\item \textbf{Realistic Vision v5}~\cite{rv5} was trained and fine-tuned over Stable Diffusion 1.5~\cite{sd15}, with a focus on generating photo-realistic images. However, because Stable Diffusion 1.5 was trained using 512$\times$512 images, Realistic Vision 5 output is also limited to 512$\times$512 pixels. 
Generating images at a resolution of 1024$\times$1024 pixels with a model originally trained for 512$\times$512 pixels often leads to the emergence of undesirable artifacts, as the diffusion models tend to replicate the quantity of subjects. 
But for an equitable comparison with other images in our study, we upscaled Realistic Vision v5 images with \textit{Highres.fix} and ScuNET-PSNR upscaler to 1024$\times$1024 pixels. 
\textit{Highres.fix} is a generation technique where the first half of sampling iterations are done at native resolution, and then the later half at the targeted upscaled resolution. This can add more accurate details to the upscaled images while the generation prompt is still in context, without the aforementioned artifacts being introduced.

\end{itemize}

\section{Pre-Survey Details and Results}
\label{appendix:presurvey-details}

\begin{figure}[H]
  \centering
  \fbox{%
    \begin{minipage}{\linewidth}
      \textbf{Age:}
      \begin{itemize}[noitemsep,label=$\square$]
        \item 18-24
        \item 25-34
        \item 35-44
        \item 45-54
        \item 55-64
        \item 65-74
        \item 75+
      \end{itemize}
      \textbf{Gender:}
      \begin{itemize}[noitemsep,label=$\square$]
        \item Male
        \item Female
        \item Other
      \end{itemize}
      \textbf{[For University Participants] Major/Department:}\\
      \_\_\_\_\_\_\_\_\_\_\_\_\_\_\_\\[1ex]
      \textbf{[For MTurk Participants] Occupation:}
      \begin{itemize}[noitemsep,label=$\square$]
        \item Professional/Managerial
        \item Skilled Trades
        \item Service Industry
        \item Sales and Marketing
        \item Healthcare and Medical
        \item Arts and Entertainment
        \item Education and Academia
        \item Manufacturing and Production
        \item Other
      \end{itemize}
      \textbf{Country:} \\\_\_\_\_\_\_\_\_\_\_\_\_\_\_\_\\[1ex]
      \textbf{Familiarity with AI generation tools:}\\
      \textbf{Text Generation Tools:}
      \begin{itemize}[noitemsep,label=$\square$]
        \item Not at All Familiar
        \item Slightly Familiar
        \item Somewhat Familiar
        \item Very Familiar
      \end{itemize}
      \textbf{Image Generation Tools:}
      \begin{itemize}[noitemsep,label=$\square$]
        \item Not at All Familiar
        \item Slightly Familiar
        \item Somewhat Familiar
        \item Very Familiar
      \end{itemize}
      \textbf{Audio Generation Tools:}
      \begin{itemize}[noitemsep,label=$\square$]
        \item Not at All Familiar
        \item Slightly Familiar
        \item Somewhat Familiar
        \item Very Familiar
      \end{itemize}
      \textbf{Video Generation Tools:}
      \begin{itemize}[noitemsep,label=$\square$]
        \item Not at All Familiar
        \item Slightly Familiar
        \item Somewhat Familiar
        \item Very Familiar
      \end{itemize}
    \end{minipage}%
  }
  \caption{Demographic survey form.}
  \label{fig:demo-surveyform}
\end{figure}

\begin{figure}[H]
\centering
\begin{subfigure}[b]{0.49\linewidth}
    \centering
    \includegraphics[width=\textwidth]{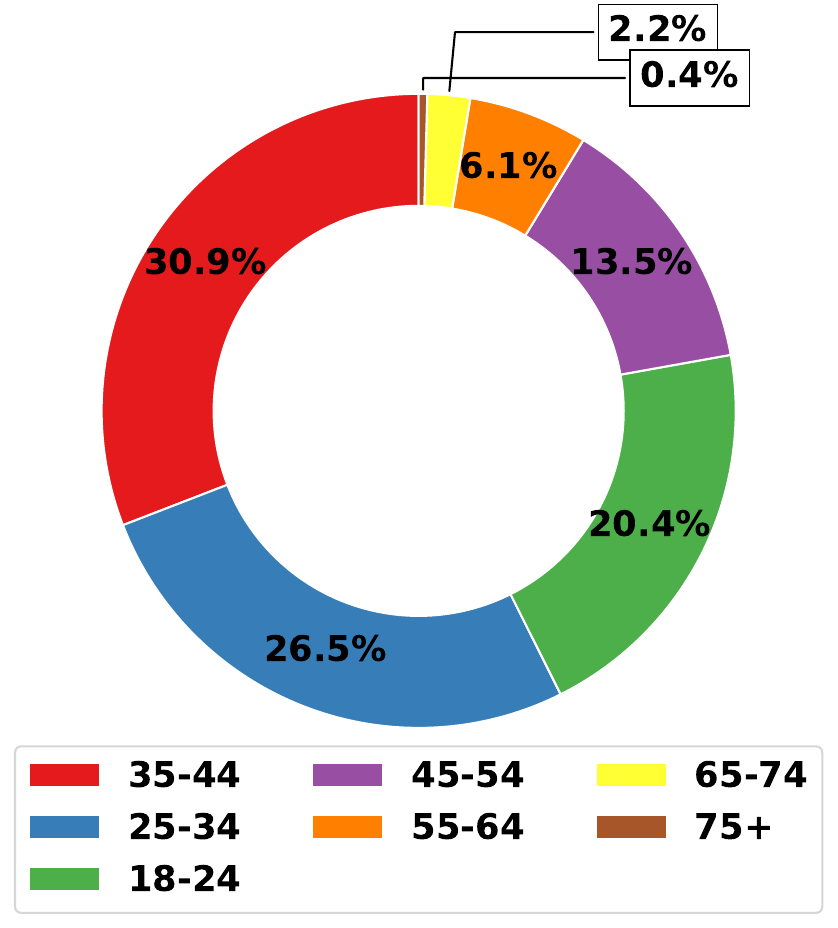}
    \caption{Age distribution.}
    \label{fig:age-distribution}
\end{subfigure}
\hfill
\begin{subfigure}[b]{0.49\linewidth}
    \centering
    \includegraphics[width=\textwidth]{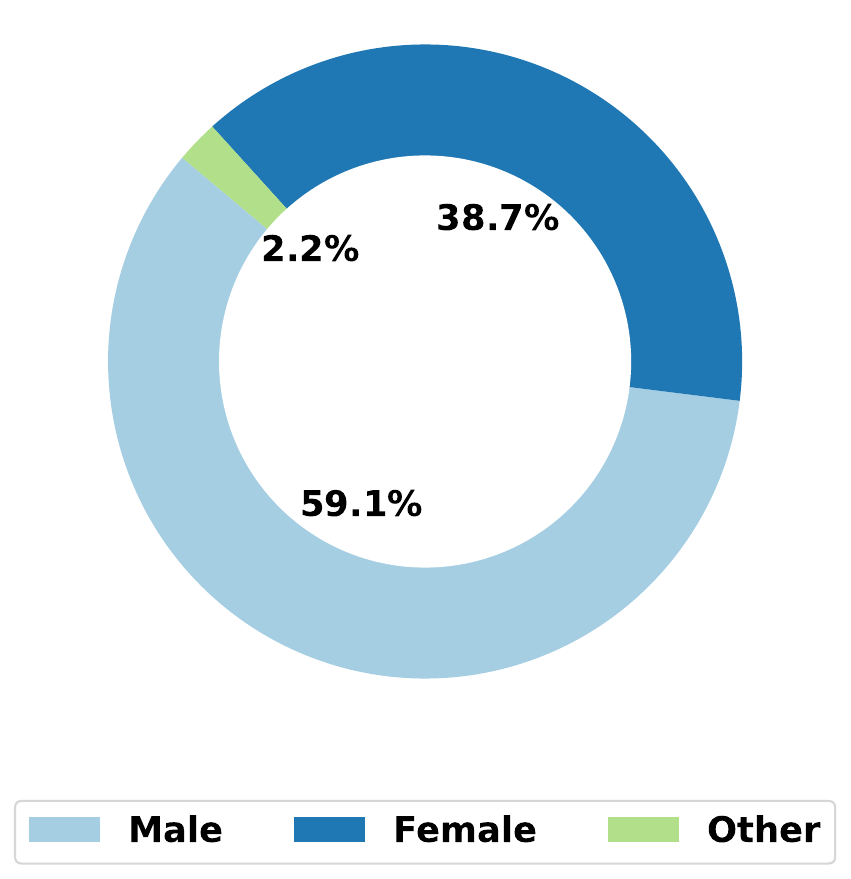}
    \caption{Gender distribution.}
    \label{fig:gender-distribution}
\end{subfigure}
\caption{Demographic distribution of survey participants.}
\label{fig:demographics}
\end{figure}

\section{Details of the Metrics}
\label{appendix:metrics}

\subsection{MSQ Scores (Survey Part~I)} 
\label{sec:metrics-msq}

The scoring mechanism for evaluating multiple select questions (MSQs) in Part~I with one or more correct answers on subject and modifiers pertaining to each displayed image is formalized through the following formula. The formula aims to give an MSQ score from 0 to 2 for each question, and is commonly used by learning management systems such as Canvas\footnote{https://www.instructure.com/canvas}. This metric serves to give an initial baseline for human prompt inference, as it only requires user to select the correct subject and modifiers from a set of given options. %
{\scriptsize
\[
MSQ Score = \max \Bigl\{0, \min \Bigl\{T, \left(T \cdot \mathbf{1}_{\{S_c=C \text{ and } I=0\}}\right) + \frac{-0.1 \times I}{C} \Bigr\} \Bigr\}
\]
}

where:
\begin{itemize}[leftmargin=*]
    \item $T = 2$ represents the total available points for a question, 1 point for correct subject selection, 1 point for correct modifier(s) selection. %
    \item $S_c$ is the number of correctly selected options by the participant.
    \item $C$ is the total number of correct options for a question.
    \item $I$ is the number of incorrect options selected by the participant.
    \item $\mathbf{1}_{\{S_c=C \text{ and } I=0\}}$ is an indicator function that equals $1$ if the participant selects all correct options without selecting any incorrect ones ($S_c=C$ and $I=0$), and $0$ otherwise.
\end{itemize}

This formula integrates the conditions for awarding full points and applying penalties for incorrect selections into a singular expression. The use of the \emph{max} and \emph{min} functions ensures that the final score remains within the acceptable range of $0$ to $T = 2$ points. The indicator function facilitates the condition under which full points are awarded, while the penalty for incorrect selections is universally applied but is effectively neutralized when full points are granted due to correct selection criteria being met. 
This scoring framework was devised to offer a balanced assessment of participant knowledge and decision-making skills, reflecting both the breadth of correct understanding and the penalties for inaccuracies. %

\subsection{Image Hash}
\begin{figure}[b]
    \centering
    \begin{subfigure}[b]{0.24\linewidth}
        \centering
        \includegraphics[width=\textwidth]{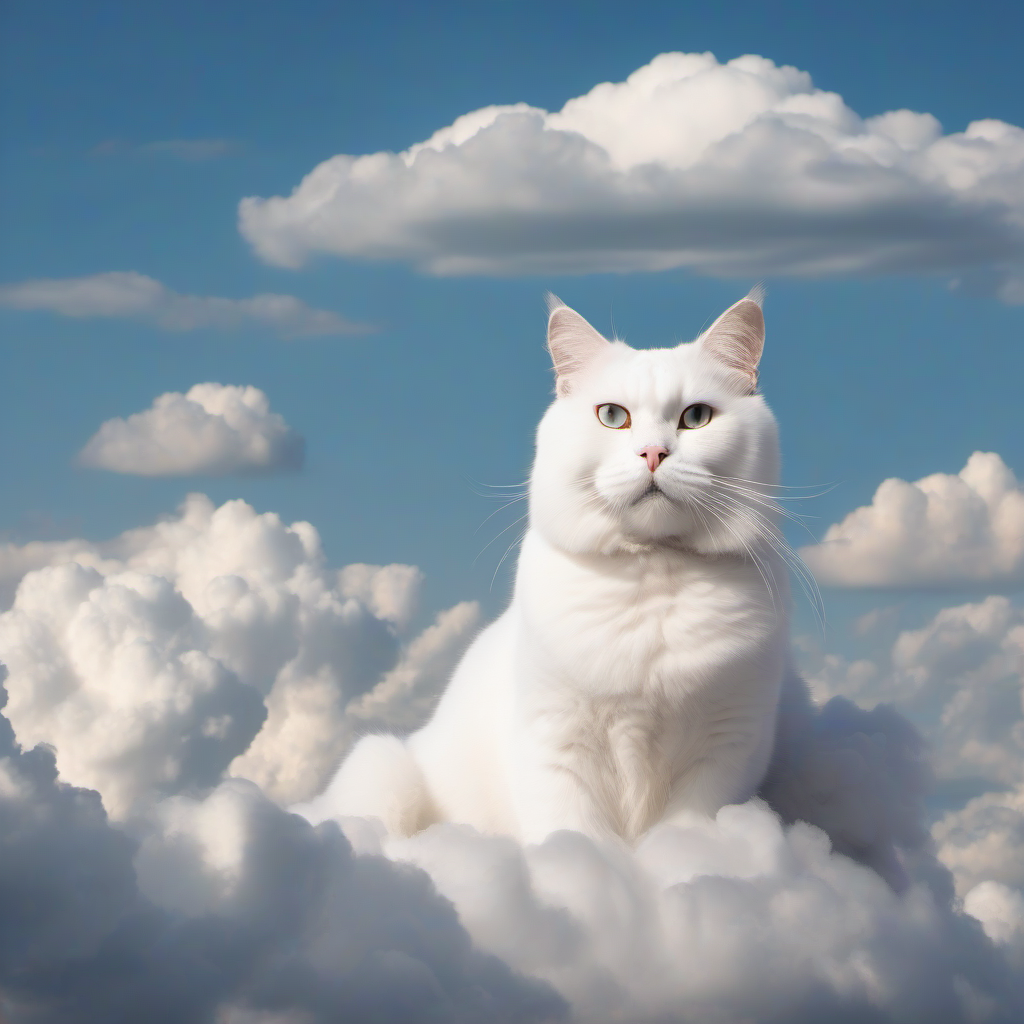}
        \caption{}
        \label{fig:cat-a}
    \end{subfigure}
    \hfill
    \begin{subfigure}[b]{0.24\linewidth}
        \centering
        \includegraphics[width=\textwidth]{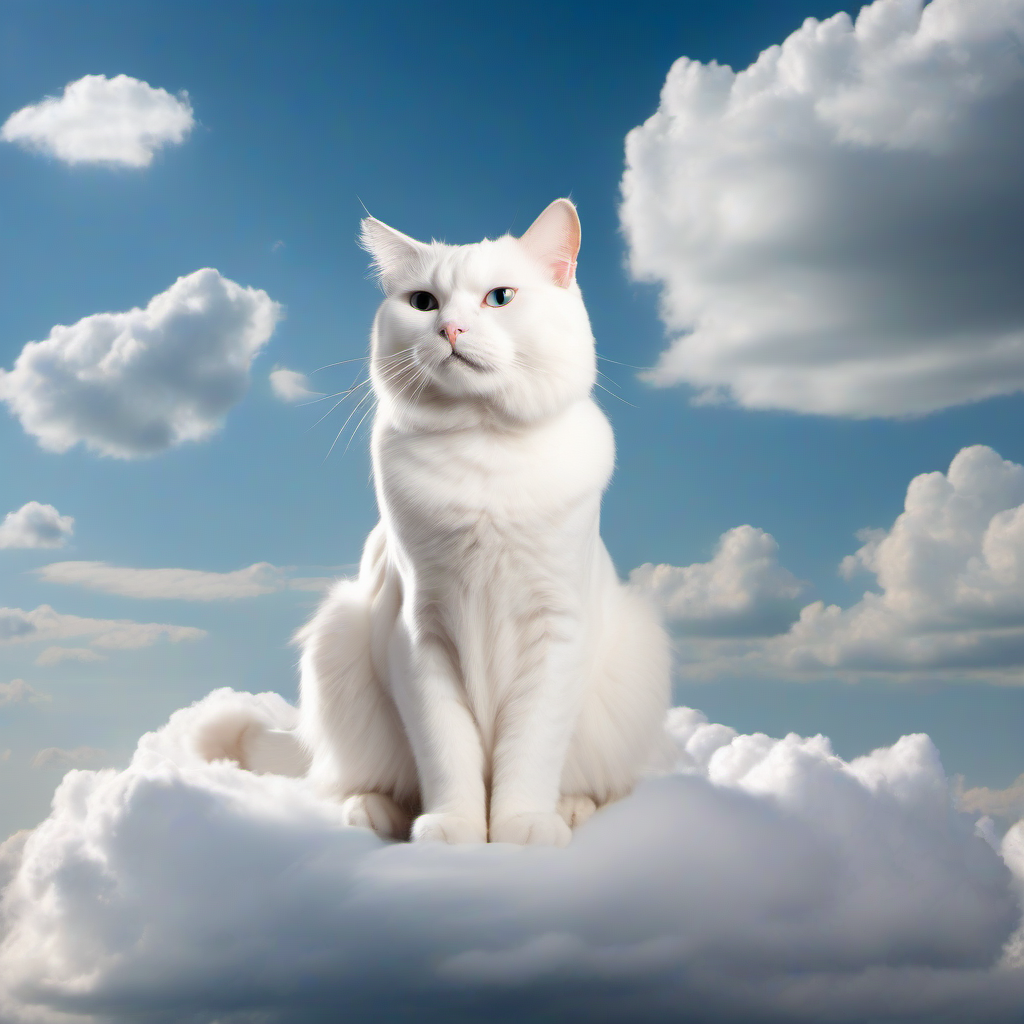}
        \caption{}
        \label{fig:cat-b}
    \end{subfigure}
     \hfill
    \begin{subfigure}[b]{0.24\linewidth}
        \centering
        \includegraphics[width=\textwidth]{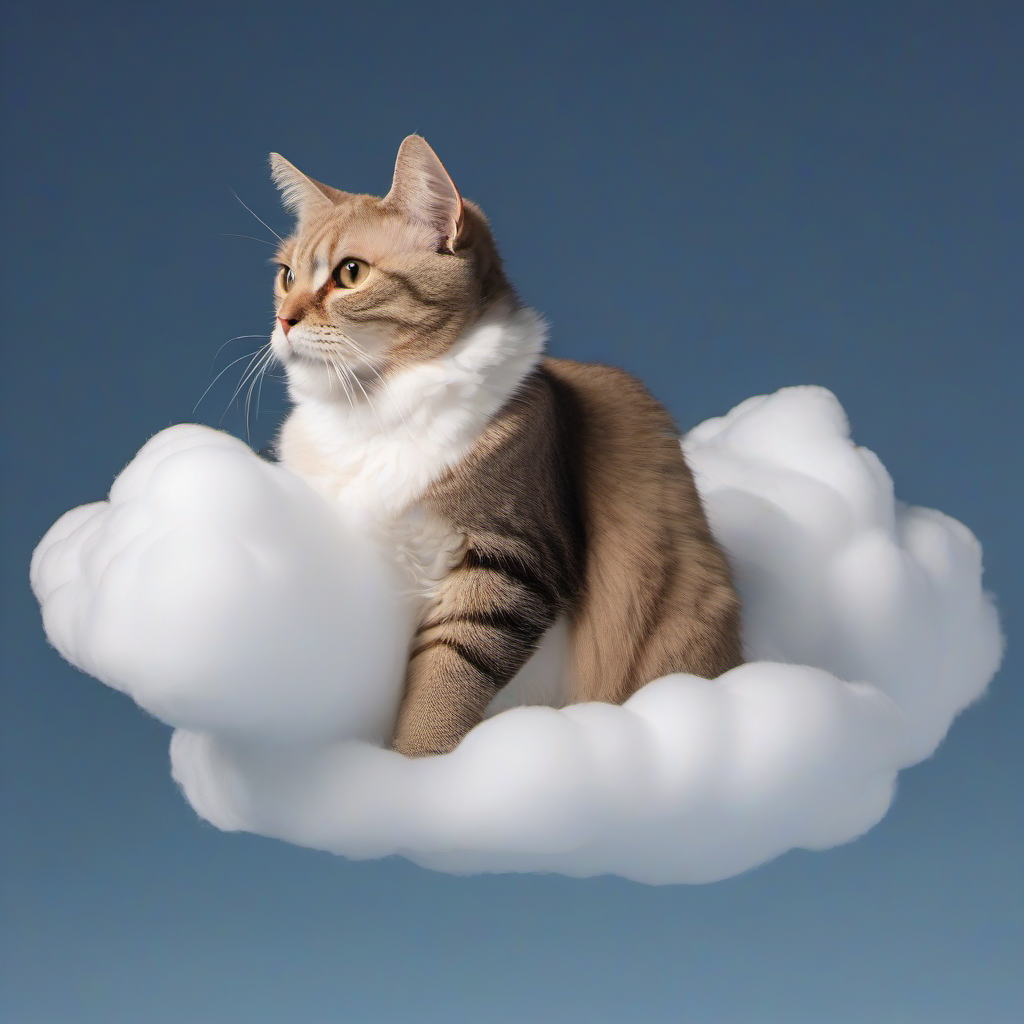}
        \caption{}
        \label{fig:cat-c}
    \end{subfigure}
     \hfill
    \begin{subfigure}[b]{0.24\linewidth}
        \centering
        \includegraphics[width=\textwidth]{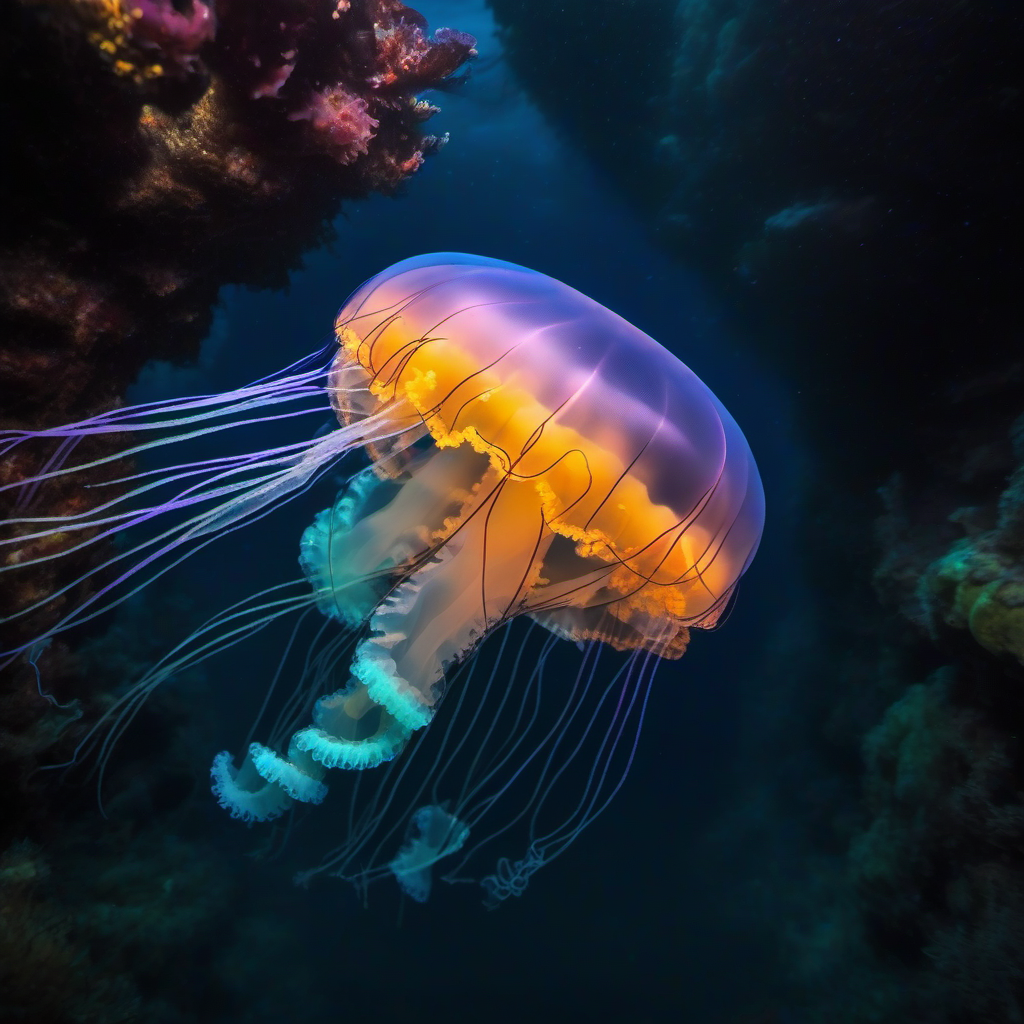}
        \caption{}
        \label{fig:jellyfish-a}
    \end{subfigure}
    \caption{SDXL generations using the same prompt ``\textit{photo of a huge white cat sitting on a cloud in the sky}'' for images (a) and (b), a similar but less specific prompt ``\textit{cat on cloud}'' for image (c), and an entirely different prompt ``\textit{a glowing jellyfish underwater, breathtaking}'' for image (d).}
    \label{fig:cats-jellyfish}
\end{figure}

The \texttt{imagehash} library in Python\footnote{\url{https://pypi.org/project/ImageHash/}} offers a compelling approach for measuring prompt inference accuracy, specifically by comparing the original AI-generated art with an art generated from an inferred prompt. \texttt{imagehash} generates perceptual hashes of images, where images that are similar result in hashes that are closely aligned. The similarity between these hashes is measured using the Hamming distance, a measure for comparing two binary data strings. By calculating the Hamming distance between the hashes of two images, we can quantitatively assess their similarity. A lower image hash score (difference) implies the two images are similar, and vice versa. For example, images in \cref{fig:cat-a} and \cref{fig:cat-b} have an image hash score of 26, where they were generated from the same prompt. In contrast, \cref{fig:cat-b} and \cref{fig:cat-c} have a slightly higher image hash score of 28 when generated from a slightly different prompt, and \cref{fig:cat-b} and \cref{fig:jellyfish-a} have a significantly higher image hash score of 34 as they were generated from very different prompts.

\subsection{Perceptual Similarity}
Perceptual similarity metric~\cite{zhang2018perceptual} captures the degree to which two images are perceived to be similar by human observers. This approach goes beyond traditional pixel-based comparisons, which often fail to capture the nuances of human visual perception. Instead, perceptual similarity considers factors like texture, color distribution, structural elements, and contextual information, which are more aligned with how humans process visual information. This method can provide a more accurate and meaningful measure of similarity between the original AI-generated art with an art generated from an inferred prompt, particularly in applications where the visual impression of an image is more important than its exact pixel composition. Similar to image hash, a lower perceptual similarity implies the two images are similar, and vice versa. For example, images in \cref{fig:cat-a} and \cref{fig:cat-b} have perceptual similarity score of 0.526, where they were generated from the same prompt. In contrast, \cref{fig:cat-a} and \cref{fig:cat-c} have a slightly higher perceptual similarity score of 0.585 when generated from a slightly different prompt, and \cref{fig:cat-a} and \cref{fig:jellyfish-a} have a significantly higher perceptual similarity score of 0.793 as they were generated from very different prompts.

\subsection{Image Embedding Similarity (CLIP Score)} 
The CLIP score between an image and a prompt (text) measures the cosine similarity between their embeddings~\cite{wang2023exploring}. A higher score indicates greater relevance or similarity, as perceived by the CLIP model, between the text and the image. This score is particularly useful in our objective to determine the accuracy or relevance of an image in relation to the textual input that was inferred to be used in its creation.
We employ two state-of-the-art CLIP models in our evaluation, OpenAI's ViT-L/14 Transformer\footnote{\url{https://huggingface.co/openai/clip-vit-large-patch14}} (L14) and ViT-B/32 Transformer\footnote{\url{https://huggingface.co/openai/clip-vit-base-patch32}} (B32).
More specifically, we utilize SentenceTransformers~\cite{reimers2019sentence} wrapped L14 and B32 models for simultaneously calculating both the CLIP score and semantic similarity when the embeddings are generated, which is described next.
Unlike image hash and perceptual similarity, a higher CLIP score implies the two images are similar, and vice versa. For example, images in \cref{fig:cat-a} and \cref{fig:cat-b} have a CLIP score of 0.970, where they were generated from the same prompt. In contrast, \cref{fig:cat-a} and \cref{fig:cat-c} have a lower CLIP score of 0.923 when generated from a slightly different prompt, and \cref{fig:cat-a} and \cref{fig:jellyfish-a} have a significantly lower CLIP score of 0.629 as they were generated from very different prompts.

\subsection{Text Embedding Similarity (Semantic Similarity)} 
In \texttt{txt2img} generation, where prompts are used to guide the creation of visual content, semantic text similarity is also an important metric in evaluating how accurately an inference deduces or reconstructed the original text prompt from the generated images.
We use SentenceTransformers~\cite{reimers2019sentence}, an extension of the BERT model for efficiently producing sentence embeddings, for calculating the cosine similarity between the original prompt and participants' response prompts.
Similar to CLIP score, a higher semantic similarity implies the two prompts are similar, and vice versa. For example, the prompts ``\textit{photo of a huge white cat sitting on a cloud in the sky}'' and ``\textit{cat on cloud}'' used in \cref{fig:cats-jellyfish} have a high semantic similarity of 0.745, as measured by the L14 model.
In contrast, the prompts ``\textit{photo of a huge white cat sitting on a cloud in the sky}'' and ``\textit{a glowing jellyfish underwater, breathtaking}'' have a semantic similarity of only 0.374.

\subsubsection{Surveyed Similarity Rating (from Survey Part~IV).} 
As outlined in \cref{sec:survey-p4}, participants in Part~IV of the study evaluated pairs consisting of an original image and a newly generated image created from a prompt provided by a previous participant. They rated these pairs on a Likert scale consisting of ``Not at All Similar'', ``Slightly Similar'', ``Somewhat Similar'', and ``Very Similar.'' This provides an additional measurement of image similarity, and thus prompt inference success, based on human interpretations.

\section{Survey Details}
\label{appendix:survey-details}

\subsection{Survey Part I Question Example} 
\label{appendix-p1}
\begin{figure}[H]
  \centering
  \includegraphics[width=0.9\linewidth]{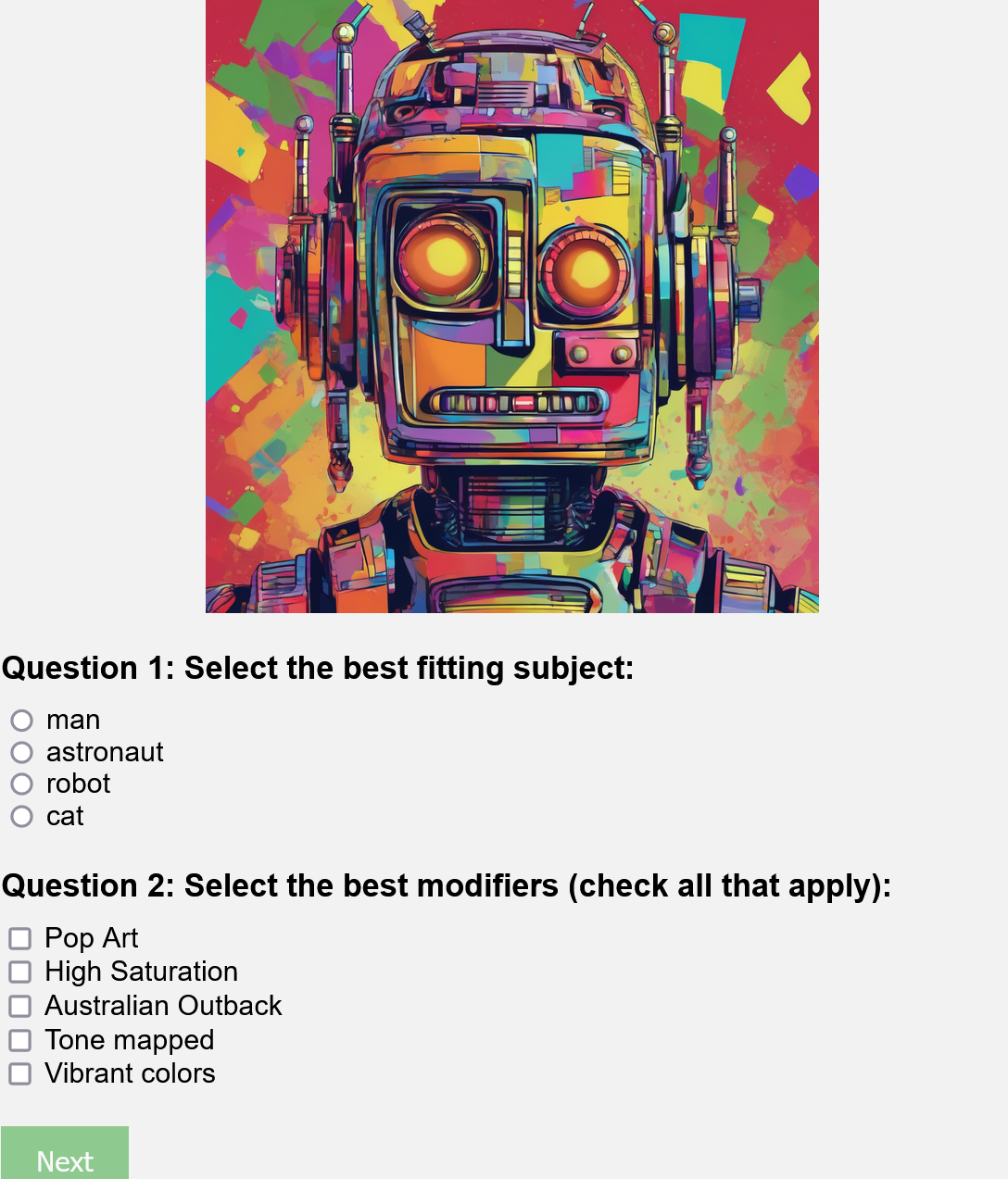}
  \caption{An example of a question for Part I. The participant must select the best fitting subject and between 1 to 5 best fitting modifiers.}
  \label{fig:p1}
\end{figure}

\subsection{Survey Part II Question Example} 
\label{appendix-p2}
\begin{figure}[H]
  \centering
  \includegraphics[width=0.9\linewidth]{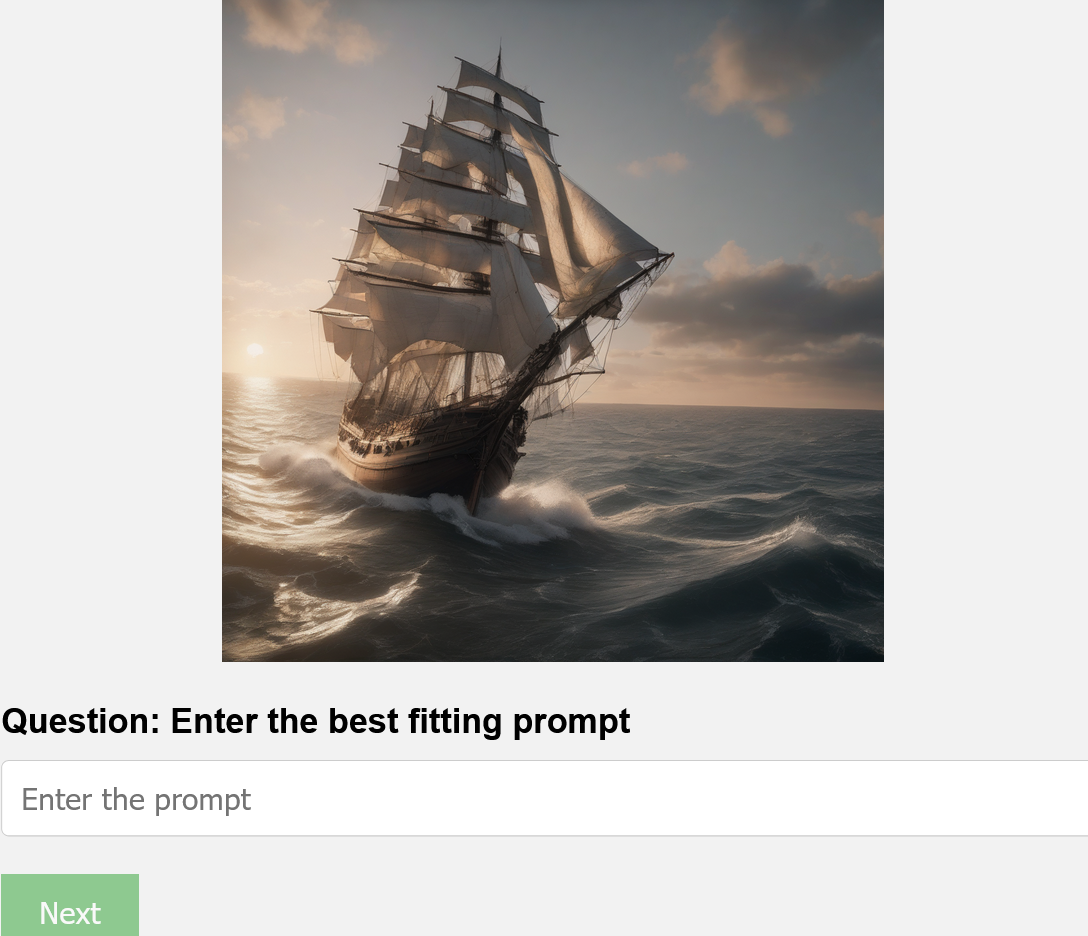}
  \caption{An example of a question for Part II. The participant must type in a prompt for the given image.}
  \label{fig:p2}
\end{figure}

\subsection{Survey Part III Question Example} 
\label{appendix-p3}
\begin{figure}[H]
  \centering
  \includegraphics[width=0.9\linewidth]{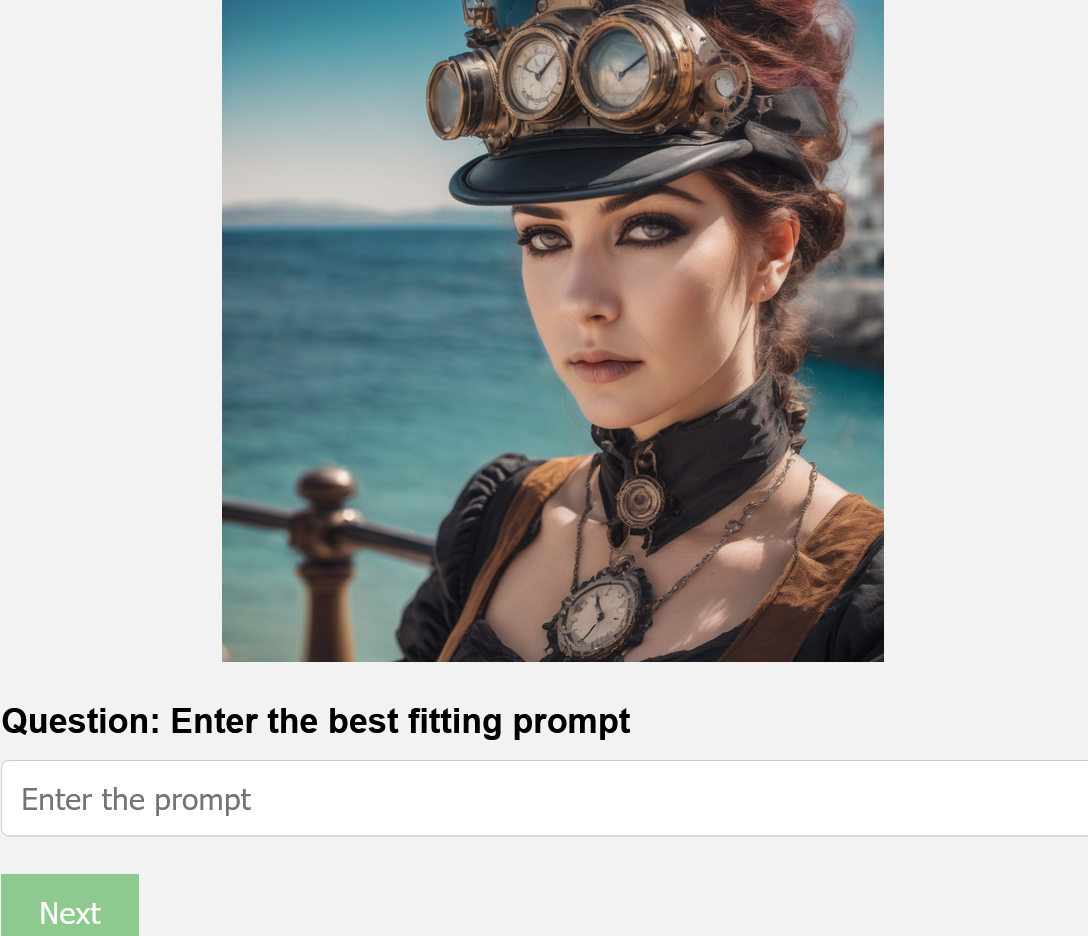}
  \caption{An example of a question for Part III. The participant must type in a prompt for the given image.}
  \label{fig:p3}
\end{figure}

\subsection{Survey Part IV Question Example}
\label{appendix-p4}
\begin{figure}[H]
\centering
\includegraphics[width=0.9\linewidth]{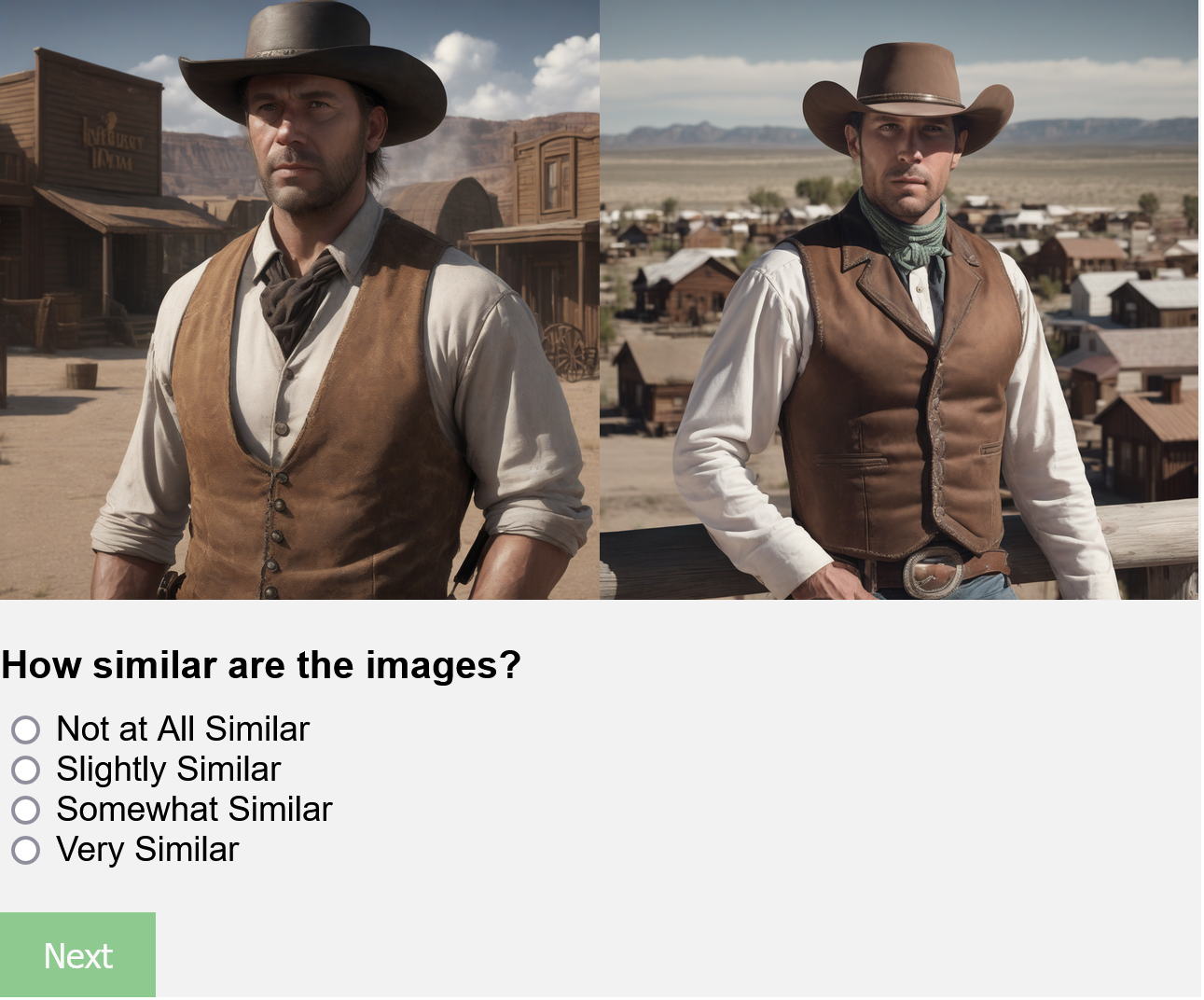}
\caption{Example of a Part~IV question, where participants rated the similarity between pairs of AI-generated images.}
\label{fig:cat-grading}
\end{figure}

\subsection{121 Controlled Dataset Modifiers and Frequency of Use} 
\label{appendix-modifiers}
abstract style (5), acrylic (3), aerial (2), african savanna (1), african savannah (1), american wild west (2), ancient chinese dynasty (4), ancient mayan (1), angry (2), arctic tundra (1), art deco (1), art deco style (1), art nouveau style (1), asymmetry (1), australian outback (2), backlit lighting (3), baroque (1), baroque style (1), bokeh lighting (3), bollywood inspired (4), bright colors (1), caribbean island (1), cartoon style (1), cel shading (4), charcoal (1), cinema render (2), cinematic lighting (1), close-up (5), cloudy lighting (2), cloudy melancholic (1), comic book (2), comic book style (3), cool colors (3), dark (3), dark colors (5), dark dramatic (1), digital art (4), digital camera (4), dramatic (4), dramatic acrylic (1), dramatic lighting (5), dreamy lighting (5), dusk lighting (5), dystopian style (2), elegant (2), engraving (2), excited (4), flat shading (1), futurist (1), futurist style (1), golden hour lighting (3), gooch shading (2), gothic european castle (6), gothic style (3), gouraud shading (2), high contrast (5), high saturation (4), high saturation sad (1), hyper real (3), japanese garden (1), joyful (1), juxtaposed (2), light-hearted (2), low contrast (4), low saturation (1), maasai village (2), macro lens (5), magical (3), mediterranean seaside (1), melancholic (1), metallic colors (7), metallic noir (1), middle eastern bazaar (2), minimalist (1), minimalist style (2), moody (1), moody lighting (1), moonlight lighting (3), moonlit (1), mysterious (1), mystical (1), mystical style (1), natural light lighting (2), octane render (1), pastel colors (2), peaceful (2), pen and ink (2), pencil sketch (5), phong shading (3), photorealistic (3), pixel art (4), polynesian island (1), pop art (1), pop art style (1), ray traced (4), realistic (5), renaissance italy (2), renaissance style (2), rustic (1), rustic style (1), sad (6), serene (3), shallow depth of field (1), sketch (2), soft lighting (1), steampunk style (3), studio (1), studio lighting (3), sunrise lighting (3), surrealist style (2), symmetry (4), telephoto lens (3), tilt-shift lens (1), tone mapped (2), unreal render (1), vibrant colors (6), victorian england (2), warm colors (7), watercolor (4), whimsical (2), wide angle (1).

\subsection{100 Controlled Dataset Prompts}
\label{appendix-prompts}
\begin{itemize}[leftmargin=*]
\item astronaut, cool colors, digital art, serene
\item astronaut, art deco
\item astronaut, ancient chinese dynasty, high contrast, pencil sketch, dramatic
\item astronaut, realistic, gothic european castle, symmetry, rustic
\item astronaut, dreamy lighting, surrealist style, peaceful, metallic colors
\item astronaut, dusk lighting, minimalist style, close-up, ancient chinese dynasty
\item astronaut, comic book style, symmetry, hyper real, sunrise lighting
\item astronaut, steampunk style, angry
\item astronaut, phong shading, dark colors
\item astronaut, dramatic lighting, renaissance style, warm colors, pen and ink
\item astronaut, excited, phong shading, high saturation, abstract style
\item astronaut, abstract style, symmetry, high saturation, photorealistic, arctic tundra
\item astronaut, gothic style, telephoto lens, dusk lighting
\item astronaut, watercolor, elegant, cool colors, natural light lighting
\item astronaut, low contrast, ray traced, backlit lighting, close-up
\item astronaut, vibrant colors, sunrise lighting, ancient chinese dynasty
\item astronaut, bollywood inspired, dusk lighting, metallic colors
\item astronaut, gothic style, dark, high contrast, studio lighting
\item astronaut, cinematic lighting, watercolor
\item cat, dreamy lighting
\item cat, dark dramatic, futurist
\item cat, cel shading, australian outback
\item cat, metallic noir, realistic, gothic european castle
\item cat, pixel art, dark colors
\item cat, dramatic lighting
\item cat, digital camera, melancholic, moonlight lighting, comic book style, ray traced
\item cat, metallic colors, golden hour lighting
\item cat, moody lighting, minimalist style, acrylic, macro lens, cel shading
\item cat, digital camera, serene, macro lens, cool colors, african savannah
\item cat, ancient chinese dynasty, bokeh lighting, gooch shading, elegant
\item cat, charcoal, tone mapped, australian outback, light-hearted, art deco style
\item cat, dreamy lighting
\item cat, pencil sketch, middle eastern bazaar, vibrant colors
\item cat, dystopian style
\item cat, magical, vibrant colors
\item cat, warm colors, dramatic lighting
\item cat, warm colors, tone mapped, engraving
\item cat, octane render, soft lighting
\item cat, telephoto lens, abstract style
\item man, baroque, comic book, minimalist
\item man, flat shading, maasai village, peaceful
\item man, bollywood inspired, acrylic, comic book, dark
\item man, dramatic, pastel colors, gothic european castle, photorealistic
\item man, realistic, dramatic lighting, excited
\item man, steampunk style, cinema render, digital art, metallic colors
\item man, digital camera, excited, renaissance italy
\item man, vibrant colors
\item man, metallic colors, whimsical, digital camera, golden hour lighting
\item man, serene
\item man, studio lighting, american wild west, unreal render
\item man, cartoon style, gouraud shading, natural light lighting, dark colors, digital art
\item man, ray traced
\item man, pop art style, phong shading
\item man, magical, watercolor, bokeh lighting, middle eastern bazaar
\item man, macro lens, high saturation sad
\item man, dark colors, caribbean island, dark, bokeh lighting, aerial
\item man, sunrise lighting
\item man, warm colors
\item man, dramatic, cinema render, warm colors, backlit lighting
\item robot, american wild west, dramatic, metallic colors
\item robot, gothic style
\item robot, moody, victorian england, sad
\item robot, pop art, high saturation
\item robot, high contrast, angry
\item robot, vibrant colors
\item robot, low contrast, realistic
\item robot, cel shading
\item robot, symmetry, warm colors, ray traced, sad, dusk lighting
\item robot, low contrast, photorealistic, magical, art nouveau style, juxtaposed
\item robot, studio lighting, surrealist style
\item robot, dystopian style
\item robot, dark colors
\item robot, dreamy lighting, pixel art, close-up, bollywood inspired, baroque style
\item robot, asymmetry, mystical style, cloudy lighting
\item robot, watercolor, futurist style, maasai village, golden hour lighting, whimsical
\item robot, sad, pixel art
\item robot, sad, low contrast, gothic european castle, sketch
\item robot, renaissance italy
\item robot, wide angle, vibrant colors, cel shading, sad
\item robot, close-up, pastel colors, ancient mayan, abstract style
\item woman, hyper real, pencil sketch, aerial
\item woman, high contrast, pencil sketch, realistic, japanese garden
\item woman, victorian england, juxtaposed, cloudy melancholic
\item woman, light-hearted, mystical
\item woman, macro lens, studio, moonlit, hyper real, gothic european castle
\item woman, rustic style, acrylic, cloudy lighting, bright colors
\item woman, dramatic lighting, telephoto lens
\item woman, steampunk style, high saturation, close-up, mediterranean seaside
\item woman, renaissance style, moonlight lighting, warm colors, dramatic acrylic
\item woman, engraving, backlit lighting, joyful
\item woman, dusk lighting, pixel art, low saturation
\item woman, gooch shading, moonlight lighting, tilt-shift lens, abstract style, mysterious
\item woman, bollywood inspired, comic book style
\item woman, polynesian island
\item woman, shallow depth of field, pen and ink, sad, metallic colors
\item woman, pencil sketch, gouraud shading
\item woman, excited, african savanna, macro lens
\item woman, gothic european castle, dreamy lighting, digital art
\item woman, sketch
\item woman, high contrast
\end{itemize}

\section{Additional Results} 
\label{appendix-additional-results}

\begin{figure}[H]
\centering
\begin{subfigure}[b]{.99\linewidth}
    \centering
    \includegraphics[width=\textwidth]{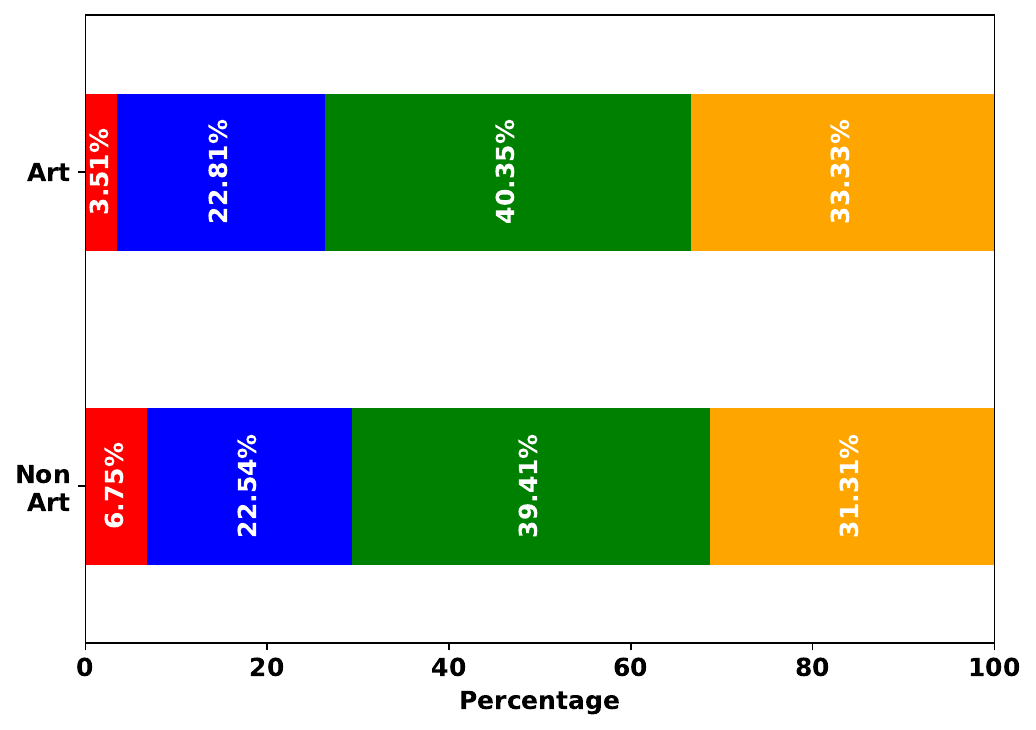}
    \label{fig:rating-art-percent}
\end{subfigure}
\hfill
\begin{subfigure}[b]{0.67\linewidth}
    \centering
    \includegraphics[width=\textwidth]{figures/part4_analysis/rating-legend-crop.pdf}
    
\end{subfigure}
\caption{Part~IV similarity rating percentage from participants with and without art background.} 
\label{fig:art-bg-rating}
\end{figure}

\begin{figure}[H]
\centering
\begin{subfigure}[b]{0.48\linewidth}
    \centering
    \includegraphics[width=\textwidth]{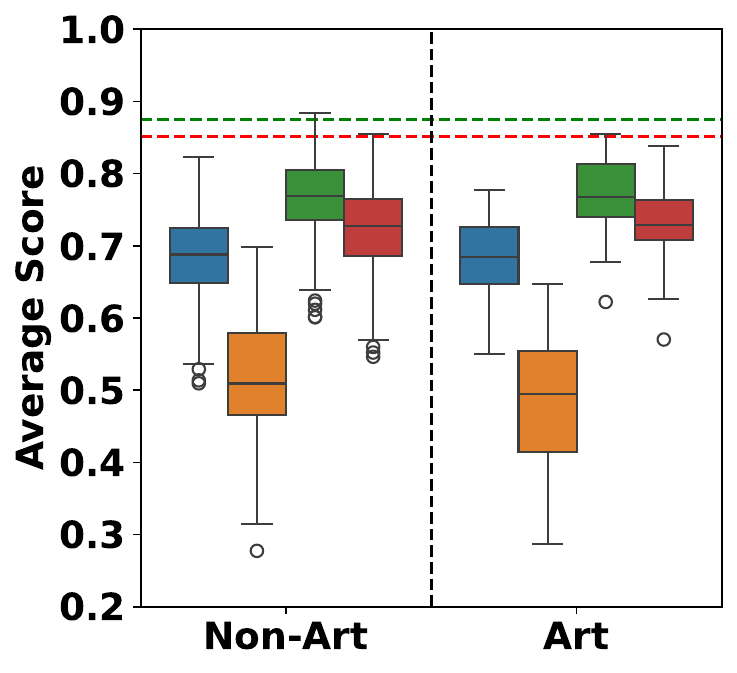}
    \caption{}
    \label{fig:art-bg-controlled}
\end{subfigure}
\hfill
\begin{subfigure}[b]{0.48\linewidth}
    \centering
    \includegraphics[width=\textwidth]{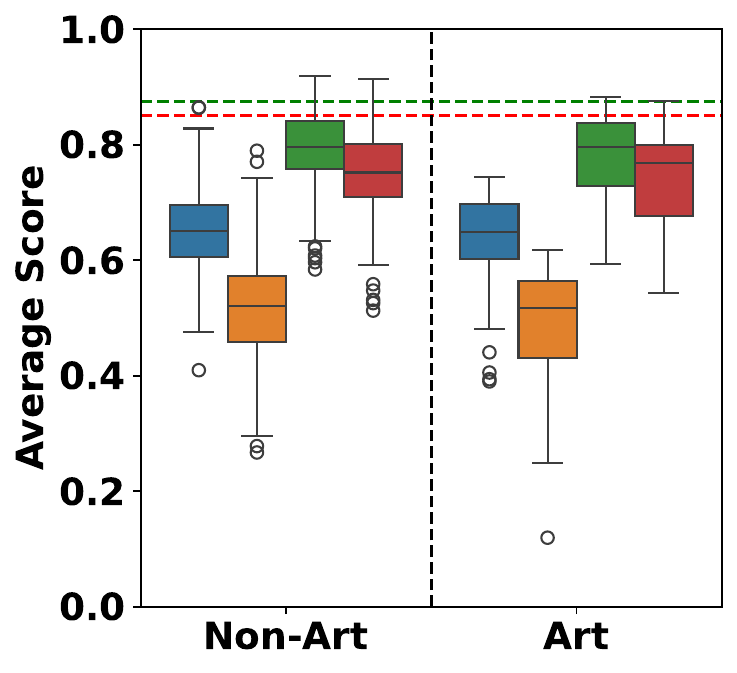}
    \caption{}
    \label{fig:art-bg-uncontrolled}
\end{subfigure}
\hfill
\begin{subfigure}[b]{0.7\linewidth}
    \centering
    \includegraphics[width=\textwidth]{figures/combined-graphs/by-models/clip-legend-crop.pdf}
    \label{fig:art-bg-legend}
\end{subfigure}
\caption{Semantic similarity and CLIP score for participants with and without art background, in (a) controlled dataset, and (b) uncontrolled dataset.}
\label{fig:art-bg-clip}
\end{figure}

\begin{figure}[H]
\centering
\begin{subfigure}[b]{0.48\linewidth}
    \centering
    \includegraphics[width=\textwidth]{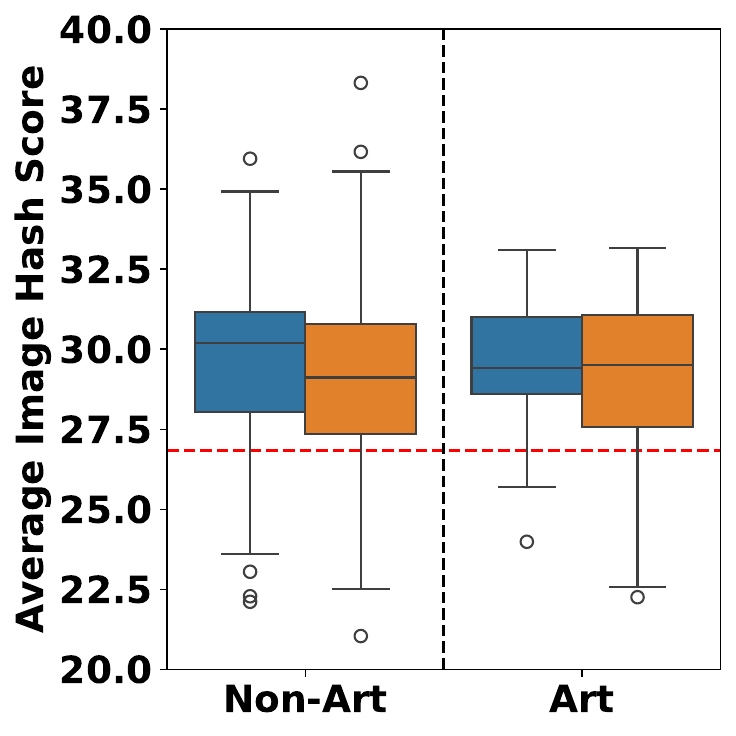}
    \caption{}
    \label{fig:art-bg-hash}
\end{subfigure}
\hfill
\begin{subfigure}[b]{0.48\linewidth}
    \centering
    \includegraphics[width=\textwidth]{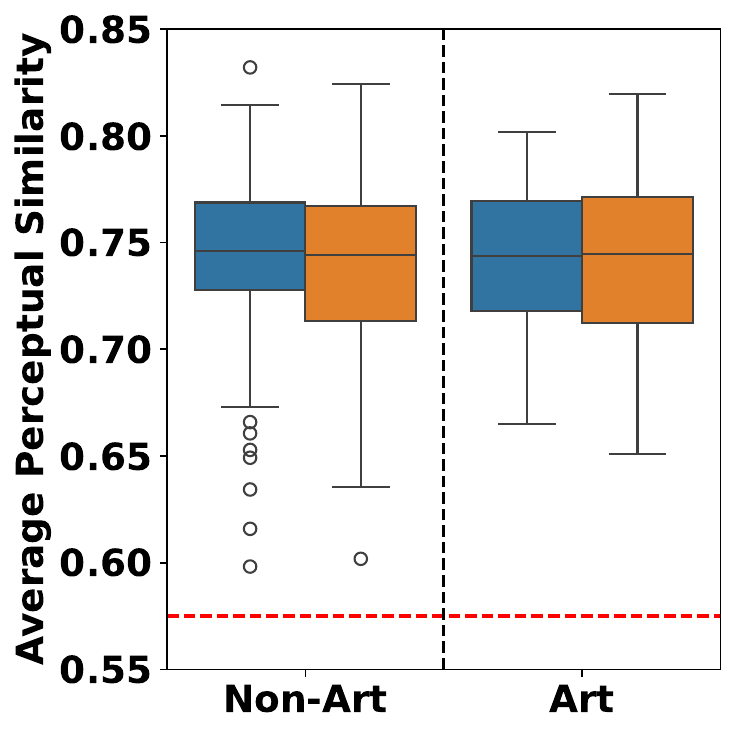}
    \caption{}
    \label{fig:art-bg-ps}
\end{subfigure}
\hfill
\begin{subfigure}[b]{0.6\linewidth}
    \centering
    \includegraphics[width=\textwidth]{figures/combined-graphs/by-models/hash-legend-crop.pdf}
    \label{fig:art-bg-ps-legend}
\end{subfigure}
\caption{Image hash score and perceptual similarity for participants with and without art background, in (a) controlled dataset, and (b) uncontrolled dataset.}
\label{fig:art-bg-hash-ps}
\end{figure}

\begin{figure}[H]
\centering
\begin{subfigure}[b]{0.48\linewidth}
    \centering
    \includegraphics[width=\textwidth]{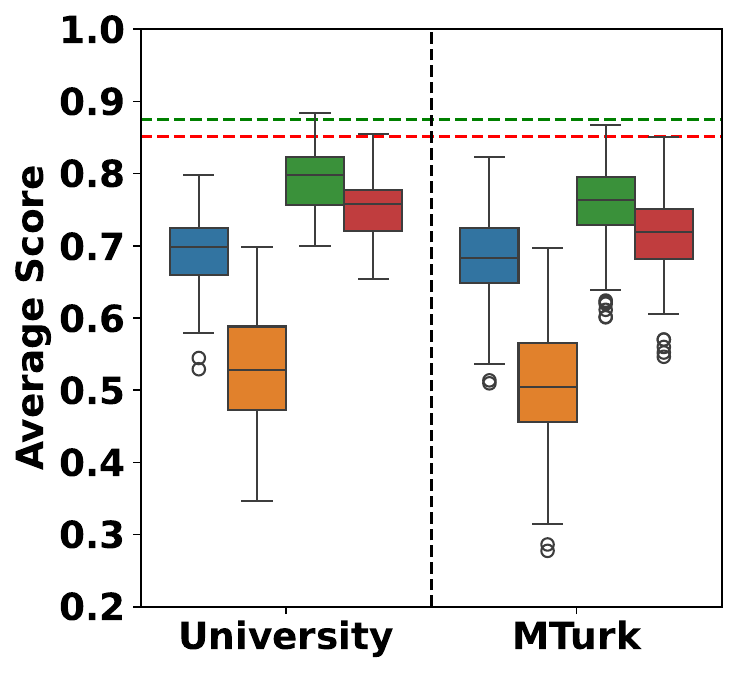}
    \caption{}
    \label{fig:mturk-controlled}
\end{subfigure}
\hfill
\begin{subfigure}[b]{0.48\linewidth}
    \centering
    \includegraphics[width=\textwidth]{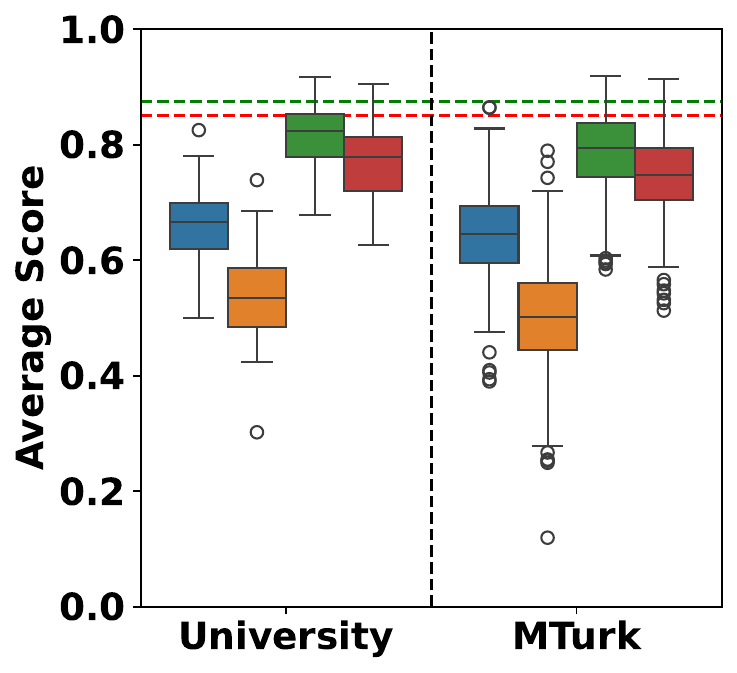}
    \caption{}
    \label{fig:mturk-uncontrolled}
\end{subfigure}
\hfill
\begin{subfigure}[b]{0.7\linewidth}
    \centering
    \includegraphics[width=\textwidth]{figures/combined-graphs/by-models/clip-legend-crop.pdf}
    
    \label{fig:mturk-legend}
\end{subfigure}
\caption{Semantic similarity and CLIP score for participants recruited on campus vs. on MTurk, in (a) controlled dataset, and (b) uncontrolled dataset.} %
\label{fig:mturk-clip}
\end{figure}

\begin{figure}[H]
\centering
\begin{subfigure}[b]{0.48\linewidth}
    \centering
    \includegraphics[width=\textwidth]{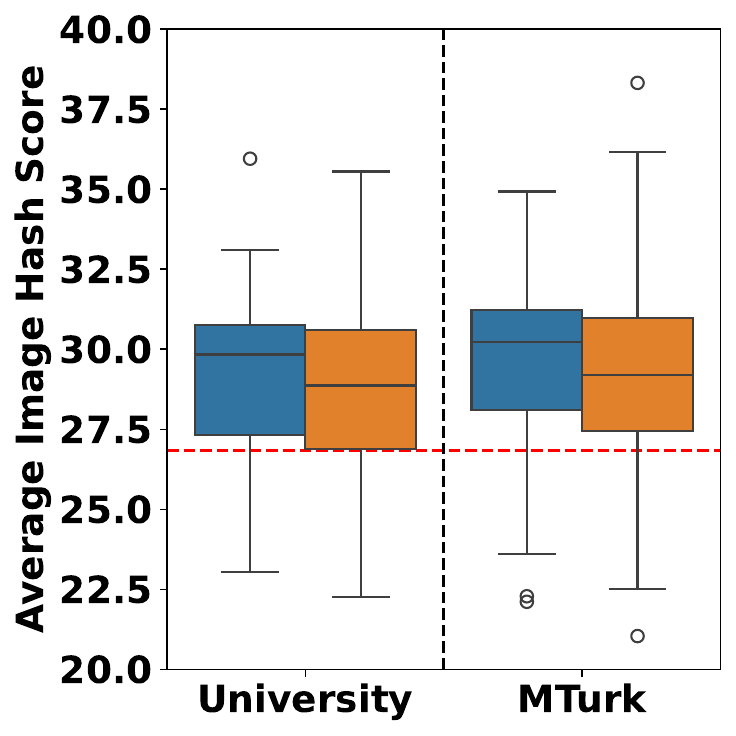}
    \label{fig:mturk-hash}
\end{subfigure}
\hfill
\begin{subfigure}[b]{0.48\linewidth}
    \centering
    \includegraphics[width=\textwidth]{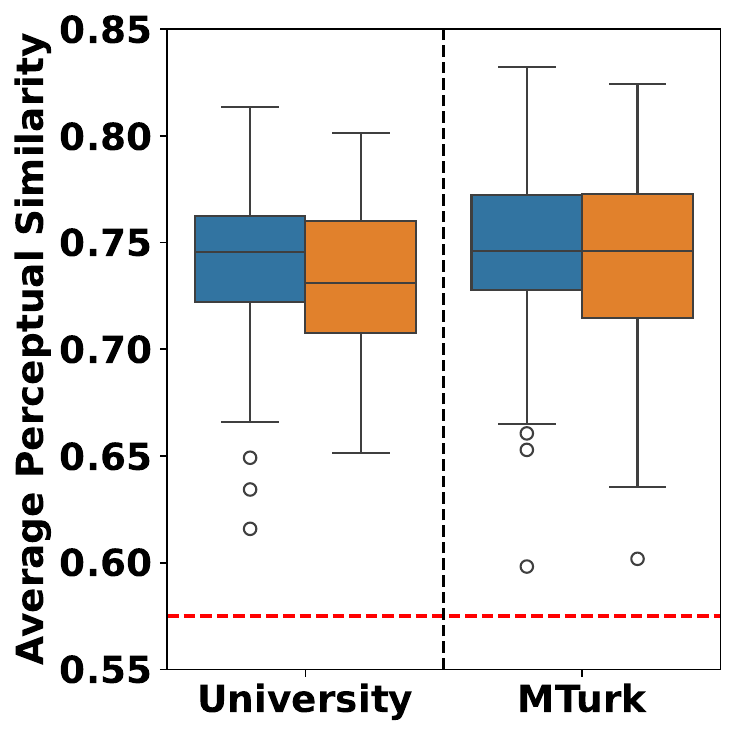}
    \label{fig:mturk-ps}
\end{subfigure}
\hfill
\begin{subfigure}[b]{0.6\linewidth}
    \centering
    \includegraphics[width=\textwidth]{figures/combined-graphs/by-models/hash-legend-crop.pdf}
    \label{fig:mturk-ps-legend}
\end{subfigure}
\caption{Image hash score and perceptual similarity for participants recruited on campus vs. on MTurk, in (a) controlled dataset, and (b) uncontrolled dataset.}
\label{fig:mturk-hash-ps}
\end{figure}

\begin{figure}[H]
\centering
\begin{subfigure}[b]{0.45\linewidth}
    \centering
    \includegraphics[width=\textwidth]{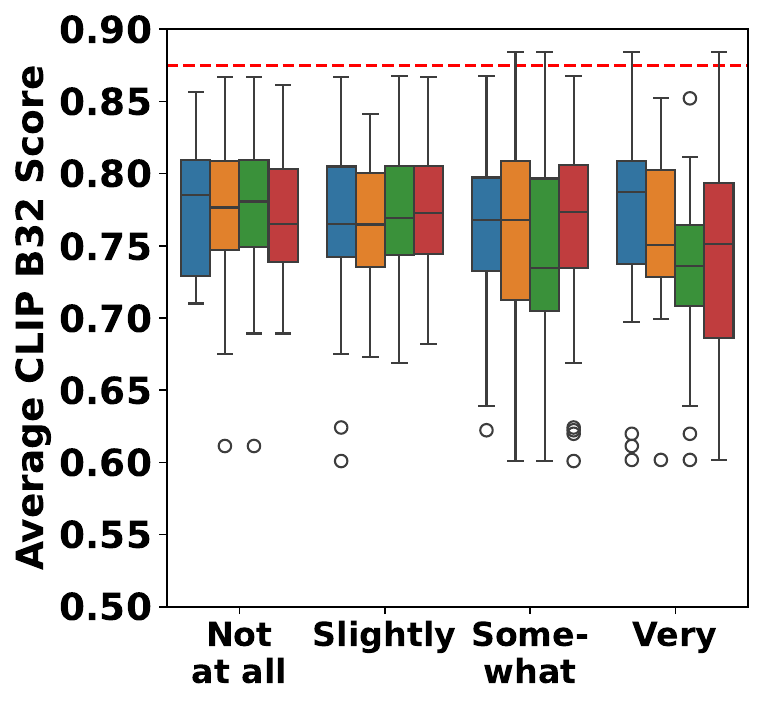}
    \caption{}
    \label{fig:controlled-tool-b32-score}
\end{subfigure}
\hfill
\begin{subfigure}[b]{0.45\linewidth}
    \centering
    \includegraphics[width=\textwidth]{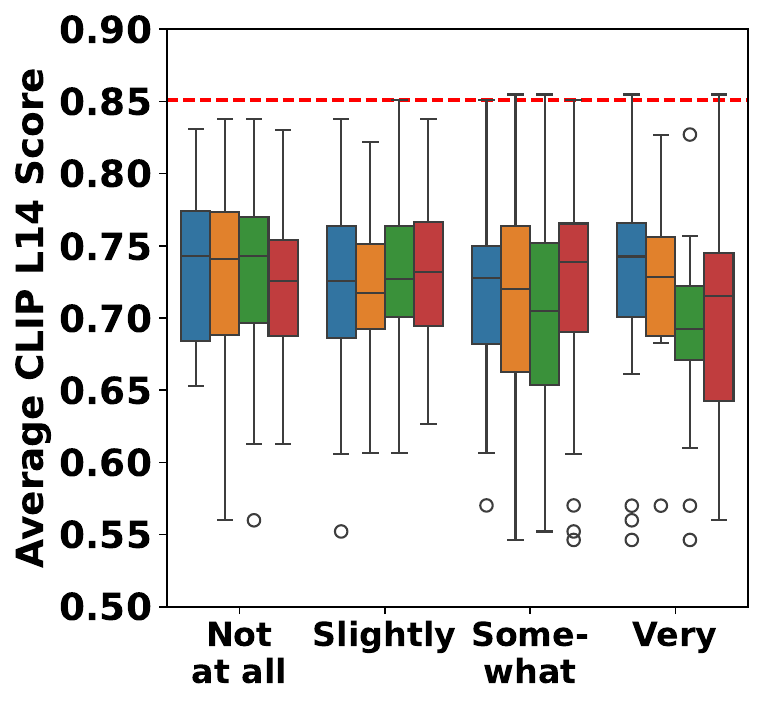}
    \caption{}
    \label{fig:controlled-tool-l14-score}
\end{subfigure}
\hfill
\begin{subfigure}[b]{0.45\linewidth}
    \centering
    \includegraphics[width=\textwidth]{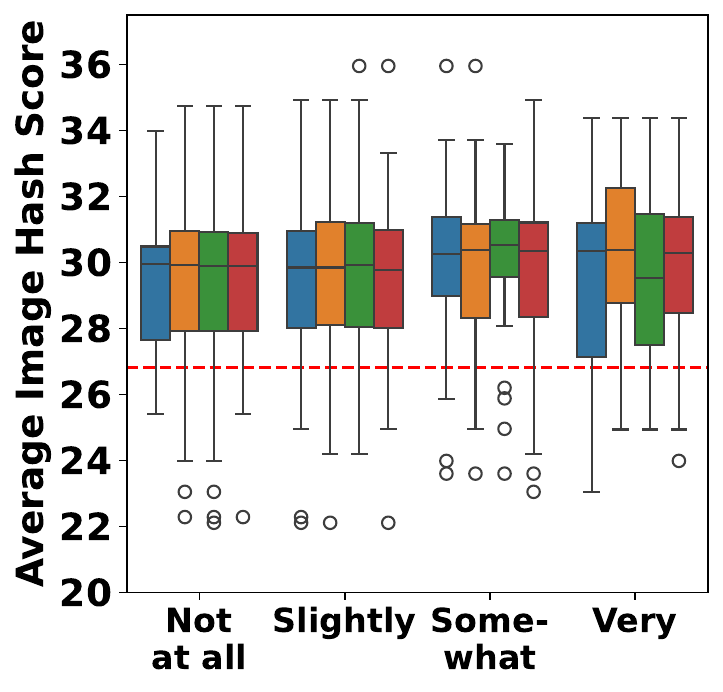}
    \caption{}
    \label{fig:controlled-hash-score}
\end{subfigure}
\hfill
\begin{subfigure}[b]{0.45\linewidth}
    \centering
    \includegraphics[width=\textwidth]{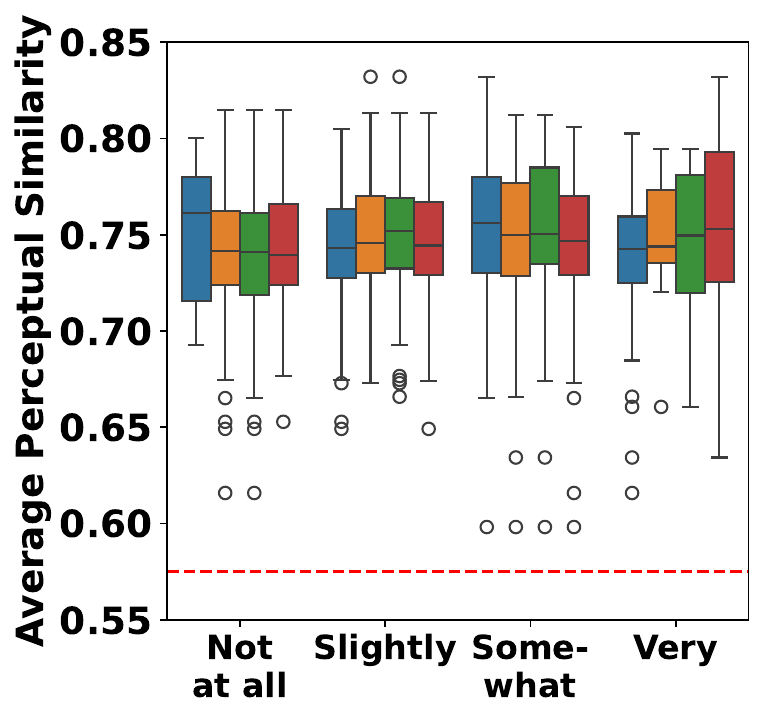}
    \caption{}
    \label{fig:controlled-tool-ps-score}
\end{subfigure}
\hfill
\begin{subfigure}[b]{0.7\linewidth}
    \centering
    \includegraphics[width=\textwidth]{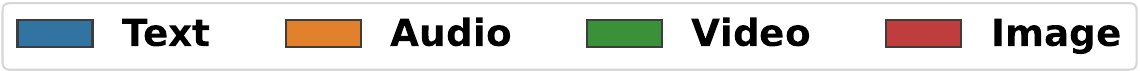}
    \label{fig:controlled-tool-legend}
\end{subfigure}
\caption{Prompt inference accuracy measured for participants with different levels of AI tools familiarity, using (a,b) B32 and L14 CLIP scores, respectively, (c) image hash, and (d) perceptual similarity, in the controlled dataset.}
\label{fig:controlled-tool-visual-score}
\end{figure}

\begin{figure}[H]
\centering
\begin{subfigure}[b]{0.45\linewidth}
    \centering
    \includegraphics[width=\textwidth]{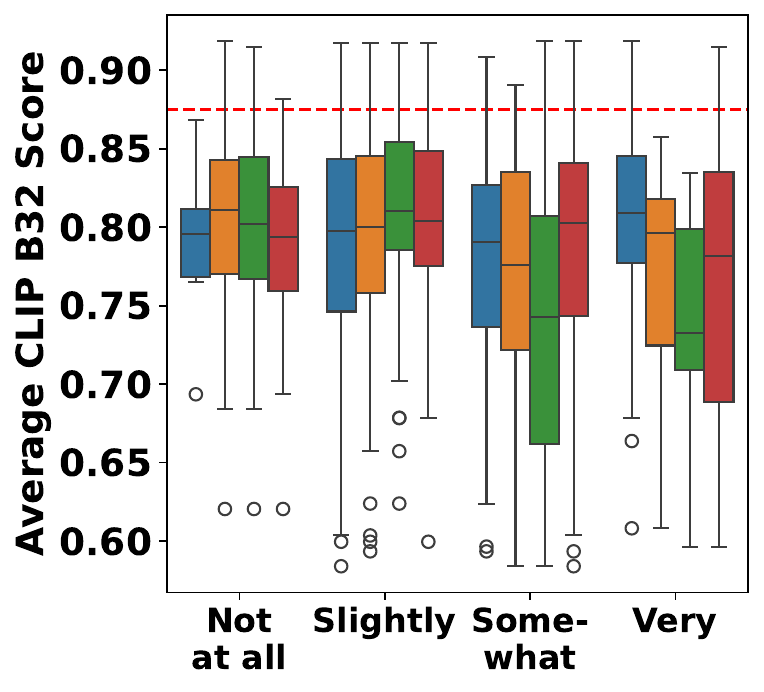}
    \caption{}
    \label{fig:uncontrolled-tool-b32-score}
\end{subfigure}
\hfill
\begin{subfigure}[b]{0.45\linewidth}
    \centering
    \includegraphics[width=\textwidth]{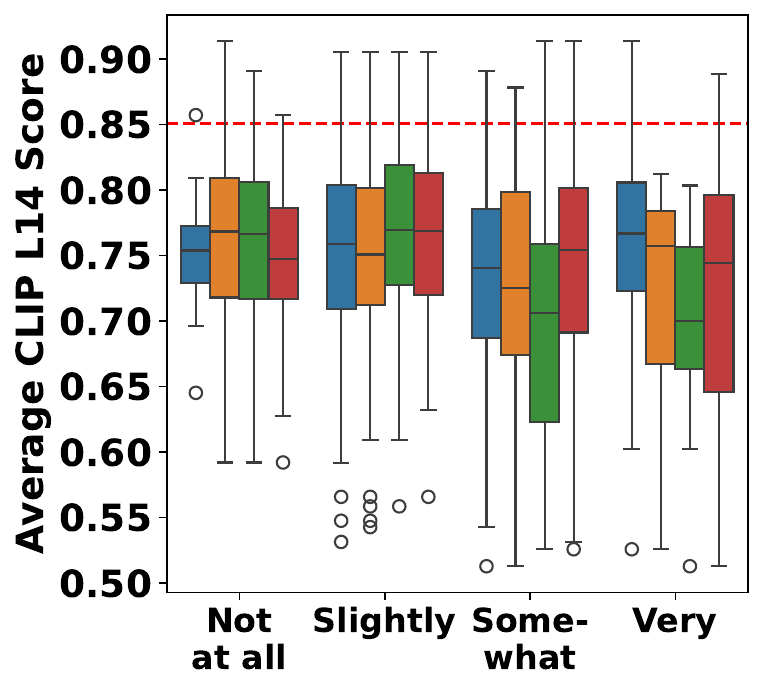}
    \caption{}
    \label{fig:uncontrolled-tool-l14-score}
\end{subfigure}
\hfill
\begin{subfigure}[b]{0.45\linewidth}
    \centering
    \includegraphics[width=\textwidth]{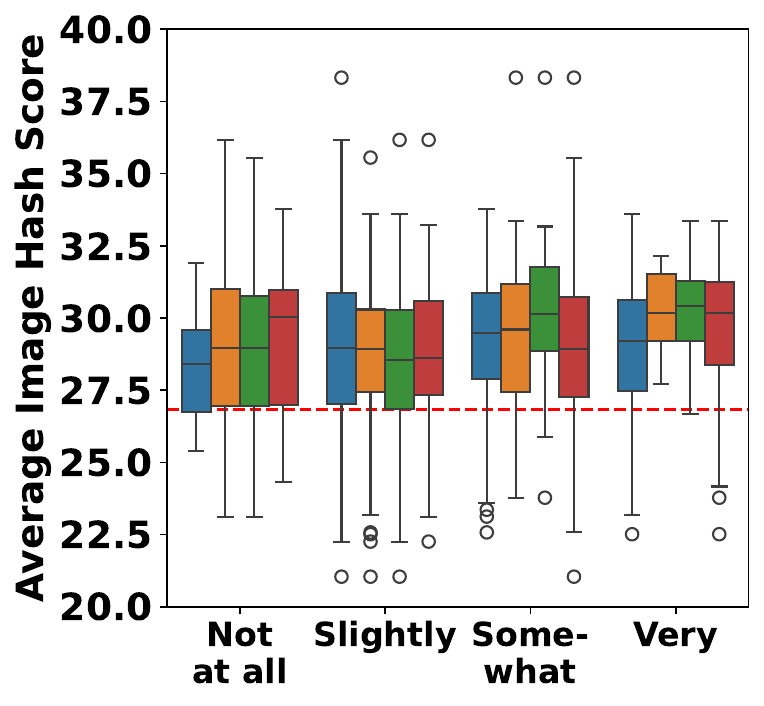}
    \caption{}
    \label{fig:uncontrolled-hash-score}
\end{subfigure}
\hfill
\begin{subfigure}[b]{0.45\linewidth}
    \centering
    \includegraphics[width=\textwidth]{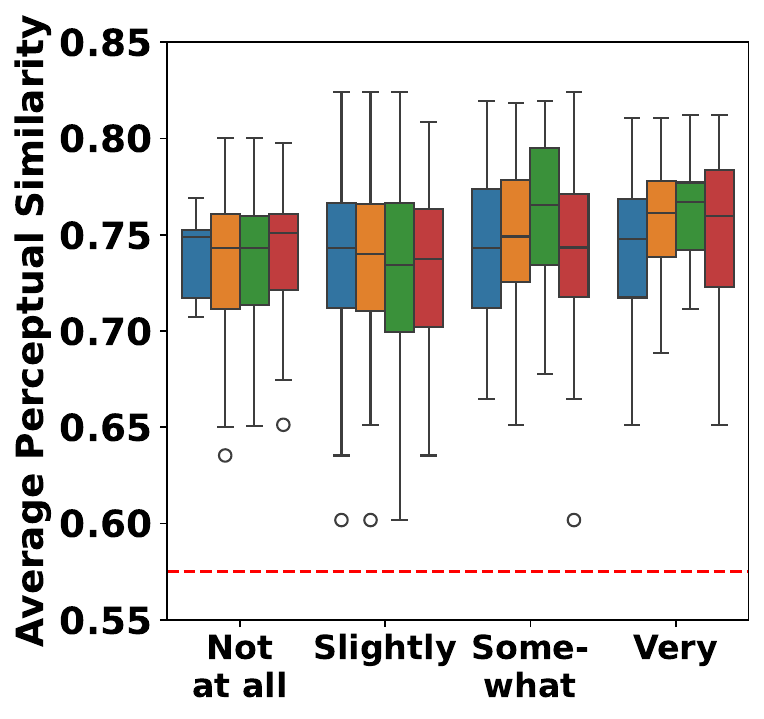}
    \caption{}
    \label{fig:uncontrolled-tool-ps-score}
\end{subfigure}
\hfill
\begin{subfigure}[b]{0.7\linewidth}
    \centering
    \includegraphics[width=\textwidth]{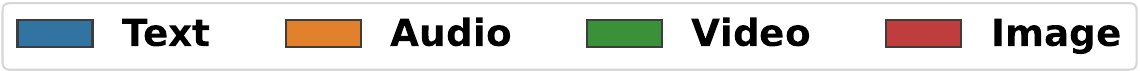}
    \label{fig:uncontrolled-tool-legend}
\end{subfigure}
\caption{Prompt inference accuracy measured for participants with different levels of AI tools familiarity, using (a,b) B32 and L14 CLIP scores, respectively, (c) image hash, and (d) perceptual similarity, in the uncontrolled dataset.}
\label{fig:uncontrolled-tool-visual-score}
\end{figure}

\end{document}